\documentclass[prc,twocolumn,nofootinbib,superscriptaddress,showpacs]{revtex4}
\usepackage{graphicx,amsmath,amssymb,bm,multirow}

\newcommand{\be}[1]{\begin{equation}\label{#1}}
\newcommand{\ee}{\end{equation}}

\newcommand{\mev}{\, \text{MeV}}

\newcommand{\bra}[1]{\langle#1|}
\newcommand{\ket}[1]{|#1\rangle}
\newcommand{\braket}[2]{\langle#1|#2\rangle}
\newcommand{\matrEL}[3]{\langle#1|#2|#3\rangle}

\newcommand{\elemA}[2]{\ensuremath{{}^{#1}}\textrm{#2}}

\newcommand{\nn}{\underline{n}}
\newcommand{\NN}{\underline{N}}
\begin{document}

\title{
Unified {\it ab initio} approach to bound and unbound states: \\
no-core shell model with continuum and its application to $^7$He}

\author{Simone Baroni}
\email[E-mail:~]{simone.baroni@ulb.ac.be}
\affiliation{Physique Nucl\'eaire Th\'eorique, Universit\'e Libre de Bruxelles, 
C.P. 229, B-1050 Bruxelles, Belgium}
\affiliation{TRIUMF, 4004 Wesbrook Mall, Vancouver BC, V6T 2A3, Canada}
\author{Petr Navr\'atil}
\email[E-mail:~]{navratil@triumf.ca}
\affiliation{TRIUMF, 4004 Wesbrook Mall, Vancouver BC, V6T 2A3, Canada}
\affiliation{Lawrence Livermore National Laboratory, P.O Box 808, L-414, Livermore, 
California 94551, USA}
\author{Sofia Quaglioni}
\email[E-mail:~]{quaglioni1@llnl.gov}
\affiliation{Lawrence Livermore National Laboratory, P.O Box 808, L-414, Livermore, 
California 94551, USA}

\begin{abstract}
We introduce a unified approach to nuclear bound and continuum states 
based on the coupling of the no-core shell model (NCSM), a bound-state technique, with the no-core shell model/resonating group method (NCSM/RGM), a nuclear scattering technique. This new {\it ab initio} method, no-core shell model with continuum (NCSMC), leads to 
convergence properties superior to either NCSM or NCSM/RGM  
while providing a balanced approach to different classes of states. In the NCSMC, the ansatz for the many-nucleon wave function includes: $i)$ a square-integrable $A$-nucleon component expanded in a complete harmonic oscillator basis; $ii)$ a binary-cluster component with 
asymptotic boundary conditions that can properly describe 
 weakly-bound states, resonances and scattering; and, 
in principle, $iii)$ a three-cluster 
component suitable for the description of, e.g., Borromean halo nuclei and reactions with final three-body states. The Schr\"odinger equation is transformed into a system of coupled-channel integral-differential equations that we solve using a modified microscopic R-matrix formalism within 
a Lagrange mesh basis. We demonstrate the usefulness of the approach by investigating the unbound $^7$He nucleus. 
\end{abstract}


\maketitle


\section{Introduction}
\label{Sec:Intro}

One of the central goals of nuclear physics is to come to a basic understanding of the structure and dynamics of nuclei, quantum many-body systems exhibiting bound states, unbound resonances, and scattering states, all of which can be strongly coupled. {\em Ab initio} ({\em i.e.}, from first principles) approaches attempt to achieve such a goal for light nuclei. 
Over the past fifteen years, 
efficient techniques such as the Green's function Monte Carlo (GFMC)~\cite{GFMC}, {\it ab initio} NCSM~\cite{Navratil:2007we}, Coupled Cluster Method (CCM)~\cite{Ha08,Hagen:2012sh,Roth:2011vt} or nuclear lattice effective field theory (EFT)~\cite{Epelbaum:2011md} have greatly advanced our understanding of bound-state properties of light nuclei starting from realistic nucleon-nucleon ($NN$) and three-nucleon ($NNN$) interactions.
On the other hand, a fully-developed fundamental theory 
able to address a large range of nuclear scattering and nuclear reaction properties is still missing,  particularly 
for processes involving more than four nucleons overall.
Better still, achieving a realistic {\em ab initio} description of light nuclei requires abandoning the ``traditional'' separated treatment of discrete states and scattering continuum in favor of a unified treatment of structural and reaction properties. 

The development of such a unified fundamental theory 
is key to refining our understanding of the underlying forces 
across the nuclear landscape: from the well-bound nuclei to 
the exotic nuclei at the boundaries of stability that have become the focus of the next generation experiments with rare-isotope beams, to the low-energy fusion reactions that represent the primary energy-generation mechanism in stars, and could potentially be used for future energy generation on earth. 

In the recent past, significant effort has been devoted to extend {\em ab initio} techniques to the treatment of dynamical processes among light nuclei~\cite{Nollett:2006su,Quaglioni:2008sm,Hagen:2012rq}. To this aim, we introduced a new many-body approach based on 
expansions over fully-antisymmetric $(A-a,a)$ binary-cluster states in the spirit of the resonating-group method (RGM)~\cite{RGM,RGM1,RGM2,RGM3,Lovas98,Hofmann08}, in which each cluster of nucleons is described within the {\em ab initio} NCSM~\cite{Navratil:2000ww}. The unknown relative-motion wave functions between pairs of clusters are obtained by solving a set of non-local integral-differential coupled-channel equations and have 
appropriate bound-state and/or scattering asymptotic behavior. 
Capable of treating bound and scattering states of light nuclei in a unified formalism starting from the fundamental inter-nucleon interactions, the NCSM/RGM approach~\cite{Quaglioni:2008sm,Quaglioni:2009mn} has been 
successfully applied 
to a wide variety of binary processes, such as nucleon-$^4$He and $n{-}^7$Li scattering~\cite{Navratil:2010jn}, 
$^7$Be($p$,$\gamma$)$^8$B capture~\cite{Navratil2011379}, $d{-}^4$He scattering~\cite{Navratil:2011ay}, $^3$H($d$,$n$)$^4$He, and $^3$He($d$,$p$)$^4$He fusion~\cite{PhysRevLett.108.042503}, and an extension to the treatment of three-cluster dynamics is under development~\cite{3bcont1,FB20proc}. At the same time, these studies have highlighted practical limitations of the approach mainly related to a non-entirely efficient convergence behavior at short-to-medium distances, as discussed in the following.  
 

Two kinds of convergence patterns have to be taken into account when performing a NCSM/RGM calculation. 
First, one has to investigate the dependence on the size of the harmonic oscillator (HO) basis 
used to expand  
the NCSM eigenstates of the clusters and 
localized components of the couplings between binary-cluster states. 
This size is characterized by $N_{\rm max}$, the maximal number of HO excitations above the lowest possible configuration of the clusters. With soft similarity-renormalization-group (SRG)~\cite{SRG,Roth_SRG,Roth_2010,Bogner_2010} evolved chiral EFT $NN$ interactions~\cite{Entem:2003ft,Machleidt:2011zz}, 
employed in most NCSM/RGM calculations, HO basis sizes with 
$N_{\rm max}\sim 10{-}14$ 
are typically sufficient to reach convergence  and computationally feasible. Second, one has to study the convergence with respect to the number of clusters' eigenstates included in the 
calculation. While including only the ground state (g.s.) of the tightly-bound $^4$He in nucleon-$^4$He scattering calculations 
already leads to a very good approximation of the $A=5$ scattering phase shifts~\cite{Navratil:2010jn}, the description of the low-energy $^7$Be($p$,$\gamma$)$^8$B capture required taking into account the lowest five eigenstates of $^7$Be~\cite{Navratil2011379}. 
The convergence with the number of clusters' eigenstates becomes even more problematic for weakly-bound clusters. Calculations with composite projectiles (the lighter of the two clusters) such as $^2$H, $^3$H, and $^3$He, show that it is essential to take into account 
the virtual breakup of these 
systems even at energies much below the breakup threshold. 
Presently, this is achieved 
by including a large number of excited pseudostates~\cite{Navratil:2011ay,PhysRevLett.108.042503,Quaglioni:2012bm} of the projectile. This in turn results in a dramatic increase of complexity of the calculations as a large number of channels are coupled. 

In this paper we present a more efficient 
approach to nuclear bound and continuum states, the no-core shell model with continuum (NCSMC). We adopt an extended model space that, in addition to the continuous binary-cluster $(A-a,a)$ NCSM/RGM states, encompasses also square-integrable NCSM eigenstates of the $A$-nucleon system. 
Such eigenstates introduce in the trial wave function 
short- and medium-range $A$-nucleon correlations that in the NCSM/RGM formalism have to be treated by including a large number of excited states of the clusters. An analogous approach was suggested already in the original RGM papers~\cite{RGM,RGM1}. The idea behind the NCSMC was first mentioned in our review paper~\cite{Navratil:2009ut} and the formalism was succinctly introduced in Ref.~\cite{Baroni:2012su}, where it was applied to study of the low-lying resonances of the exotic $^7$He nucleus using an SRG-evolved chiral EFT $NN$ potential that provides an accurate description of the $NN$ system. 
Here, we give a detailed presentation of the formalism, discuss the results published in Ref.~\cite{Baroni:2012su}, and present additional results. 

In Sec.~\ref{Sec:Formalism}, we briefly review the NCSM and NCSM/RGM approaches and then introduce in detail the NCSMC formalism. In Sec.~\ref{Sec:7He}, we apply the NCSMC to the exotic $^7$He nucleus. We discuss calculations presented in Ref.~\cite{Baroni:2012su} as well as additional results. 
Conclusions and outlook are given in Sect.~\ref{concl}. Parts of the formalism not suitable for the main text are presented in Appendix~\ref{Appendix_A}.

\section{Formalism}
\label{Sec:Formalism}

This section is dedicated to the formalism of the NCSMC theory with a particular focus on the case in which the binary-cluster portion of the basis is given by a single-nucleon projectile in relative motion with respect to an $(A-1)$-nucleon target. First, in Sec.~\ref{Subsec:NCSM}, we briefly review the NCSM, then in Sec.~\ref{Subsec:NCSMRGM} we present 
useful background and expressions 
for the NCSM/RGM formalism. Finally, in Sec.~\ref{Subsec:NCSMC} we introduce in detail the NCSMC.

\subsection{NCSM}
\label{Subsec:NCSM}

The {\it ab initio} NCSM is a 
structure technique appropriate for 
the description of bound states or for approximations of narrow resonances. Nuclei are considered as systems of $A$ non-relativistic point-like nucleons interacting 
through realistic inter-nucleon interactions, i.e., 
those that describe accurately two-nucleon and, possibly, three-nucleon systems. All nucleons are active degrees of freedom. Translational invariance as well as angular momentum and parity of the system under consideration are conserved. 
The 
many-body wave function is cast into 
an expansion over 
a complete set of antisymmetric $A$-nucleon HO basis states containing up to  
$N_{\rm max}$ 
 HO excitations above the lowest possible configuration: 
\begin{equation}\label{NCSM_wav}
 \ket{\Psi^{J^\pi T}_A} = \sum_{N=0}^{N_{\rm max}}\sum_i c_{Ni}\ket{ANiJ^\pi T}\; .
\end{equation}
Here, $N$ denotes the total number of HO excitations of all nucleons above the minimum configuration,  $J^\pi T$ are the total angular momentum, parity and isospin, and $i$ additional quantum numbers. The sum over $N$ is restricted by parity to either an even or odd sequence.
The basis is further characterized by the frequency $\Omega$ of the HO well and may depend on either Jacobi relative 
or 
single-particle coordinates. In the former case, the wave function 
does not contain the center of mass (c.m.) motion, but 
antisymmetrization is complicated. In the latter case,  
antisymmetrization is trivially achieved using Slater determinants, but the c.m.\ degrees of freedom are included in the basis. The HO basis  within the $N_{\rm max}$ truncation is the only possible one that allows an exact factorization of the c.m.\ motion for the eigenstates, even when working with 
single-particle coordinates and Slater determinants. Calculations performed with the two alternative coordinate choices are completely equivalent. 

 
Square-integrable energy eigenstates expanded over the $N_{\rm max}\hbar\Omega$ basis, $\ket{ANiJ^\pi T}$, are obtained by diagonalizing the intrinsic Hamiltonian, $\hat{H}=\hat{T}_{\rm int}+\hat{V}$, 
\begin{equation}\label{NCSM_eq}
\hat{H} \ket{A \lambda J^\pi T} = E_\lambda \ket{A \lambda J^\pi T} \; ,
\end{equation}
%
where $\hat{T}_{\rm int}$ is the internal kinetic energy operator and 
$\hat{V}$ the $NN$ or $NN{+}NNN$ interaction. Convergence of the HO expansion with increasing $N_{\rm max}$ values is accelerated by the use of effective interactions derived from the underlying potential model through either Lee-Suzuki similarity transformations in the NCSM space~\cite{NO03,Navratil:2000ww} or SRG transformations in momentum space~\cite{SRG,Roth_SRG,Roth_2010,Bogner_2010,Jurgenson:2009qs,Jurgenson:2010wy}. In this latter case, the NCSM calculations are variational. Finally, we note that with the 
HO basis sizes typically used ($N_{\rm max}{\sim}10{-}14$), the $\ket{A \lambda J^\pi T}$ eigenstates lack correct asymptotic behavior for weakly-bound states and 
always have incorrect asymptotic behavior for resonances.

\subsection{NCSM/RGM}
\label{Subsec:NCSMRGM}

In the NCSM/RGM, the ansatz of Eq.~(\ref{NCSM_wav}) for the $A$-nucleon wave function is replaced by an expansion 
over 
antisymmetrized products of binary-cluster channel states 
$\ket{\Phi_{\nu r}^{J^\pi T}}$ and wave functions of their relative motion 
\begin{eqnarray}\label{eq:formalism_10}
  \ket{\Psi^{J^\pi T}_A} & = & 
                           \sum_{\nu} \int dr \: r^2 
                               \frac{\gamma_{\nu}(r)}{r}
                               \hat{\mathcal{A}}_\nu\ket{\Phi_{\nu r}^{J^\pi T}} \; .
\end{eqnarray}
The channel states 
$\ket{\Phi_{\nu r}^{J^\pi T}}$ contain $(A-a)$- and $a$-nucleon clusters 
(with $a\leq A$) of total angular momentum, parity, isospin and additional quantum number
$I_1,\pi_1,T_1,\alpha_1$ and $I_2,\pi_2,T_2,\alpha_2$, respectively, and are characterized by the relative orbital angular momentum $\ell$ and channel spin $\vec{s}=\vec{I}_1+\vec{I}_2$:
%
\begin{align}
\label{eq:formalism_20}
	\ket{\Phi_{\nu r}^{J^\pi T}} = &
	\Big[ \left(
         \ket{A-a \; \alpha_1 I_1^{\pi_1}T_1}\ket{a \; \alpha_2 I_2^{\pi_2}T_2}
         \right)^{(sT)} \nonumber\\
         &\times Y_\ell(\hat{r}_{A-a,a})
         \Big]^{(J^{\pi}T)}  \frac{\delta(r-r_{A-a,a})}{rr_{A-a,a}} \; .
\end{align}
The channel index $\nu$ collects the quantum numbers
$\{A-a \; \alpha_1 I_1^{\pi_1}T_1; a \; \alpha_2 I_2^{\pi_2}T_2; s\ell\}$. The intercluster
relative vector $\vec{r}_{A-a,a}$ is the 
displacement between the clusters' centers of mass and
is given in terms of the single-particle coordinates 
$\vec{r}_i$ by:
\begin{equation}\label{eq:formalism_30}
  \vec{r}_{A-a,a}=r_{A-a,a}\hat{r}_{A-a,a}=\frac{1}{A-a}\sum_{i=1}^{A-a}\vec{r}_i
                                       -\frac{1}{a}\sum_{j=A-a+1}^{A}\vec{r}_j\;.
\end{equation}
The cluster wave functions depend on translationally invariant internal coordinates
and are antisymmetric under exchange of internal nucleons, while the intercluster
antisymmetrizer 
$\hat{\mathcal{A}}_\nu$ takes care of the exchange of
nucleons belonging to different clusters.

With appropriate boundary conditions imposed on the wave functions of the relative motion $\gamma_{\nu}(r)$, the expansion of Eq.~(\ref{eq:formalism_10}) is suitable for describing bound states, resonances and scattering states between clusters.
For bound states, expansions (\ref{NCSM_wav}) and (\ref{eq:formalism_10}) are equivalent, although for well-bound systems where short-range $A$-body correlations play a dominant role, the convergence of the eigenenergy would typically be more efficient within the NCSM model space defined by Eq.~(\ref{NCSM_wav}). 

The unknown relative-motion wave functions $\gamma_\nu(r)$
are determined by solving the many-body Schr\"odinger equation in the Hilbert space spanned by the basis states $\hat{\mathcal A}_{\nu}\,|\Phi^{J^\pi T}_{\nu r}\rangle$:
\begin{equation}
\sum_{\nu}\int dr \,r^2\left[{\mathcal H}^{J^\pi T}_{\nu^\prime\nu}(r^\prime, r)-E\,{\mathcal N}^{J^\pi T}_{\nu^\prime\nu}(r^\prime,r)\right] \frac{\gamma_\nu(r)}{r} = 0\,,\label{RGMeq}
\end{equation}
where 
\begin{eqnarray}
{\mathcal H}^{J^\pi T}_{\nu^\prime\nu}(r^\prime, r) &=& \left\langle\Phi^{J^\pi T}_{\nu^\prime r^\prime}\right|\hat{\mathcal A}_{\nu^\prime}\hat{H}\hat{\mathcal A}_{\nu}\left|\Phi^{J^\pi T}_{\nu r}\right\rangle\,,\label{H-kernel}\\
{\mathcal N}^{J^\pi T}_{\nu^\prime\nu}(r^\prime, r) &=& \left\langle\Phi^{J^\pi T}_{\nu^\prime r^\prime}\right|\hat{\mathcal A}_{\nu^\prime}\hat{\mathcal A}_{\nu}\left|\Phi^{J^\pi T}_{\nu r}\right\rangle\,,\label{N-kernel}
\end{eqnarray}
are the Hamiltonian and norm kernels, respectively, 
and $E$ is the total energy in the c.m.\ frame. 

When computing Eqs.~(\ref{H-kernel}) and (\ref{N-kernel}), the ``exchange'' terms of the norm kernel
arising from the non-identical permutations in $\hat{\mathcal{A}}_\nu$
as well as all localized parts of the Hamiltonian kernel
are obtained by expanding the radial dependence of the basis states
of Eq.~(\ref{eq:formalism_20}) on HO radial wave functions 
$R_{n\ell}(r)$ according to:
\begin{eqnarray}\label{eq:formalism_ff_82}
\ket{\Phi_{\nu r}^{J^\pi T  }} & = & \sum_{n \in P} R_{n\ell}(r) \ket{\Phi_{\nu n}^{J^\pi T  }} \; ,
\end{eqnarray}
where $P$ indicates the HO model space and
\begin{align}
\label{eq:formalism_84}
	\ket{\Phi_{\nu n}^{J^\pi T}} = &
	\Big[ \left(
         \ket{A-a \; \alpha_1 I_1^{\pi_1}T_1}\ket{a \; \alpha_2 I_2^{\pi_2}T_2}
         \right)^{(sT)} \nonumber\\
         &\times Y_\ell(\hat{r}_{A-a,a})
         \Big]^{(J^{\pi}T)} R_{n\ell}(r_{A-a,a})   \; .
\end{align}

Here, we remind that the $A$-nucleon microscopic Hamiltonian can be written in the form
\begin{equation}\label{eq:formalism_40}
  \hat{H}=\hat{T}_{\rm rel} + \hat{\mathcal{V}}_{\rm rel} + \hat{V}_C(r) + \hat{H}_{(A-a)} + \hat{H}_{(a)}\;,
\end{equation}
where $\hat{T}_{\rm rel}$ is the relative kinetic energy between target and projectile, $\hat{\mathcal{V}}_{\rm rel}$ includes
all the interactions between nucleons belonging to different clusters after subtraction
of the average Coulomb interaction between them $\hat{V}_C(r)$ (see \cite{Quaglioni:2009mn} for
a detailed discussion on this point), and $\hat{H}_{(A-a)}$ and $\hat{H}_{(a)}$ are the intrinsic microscopic Hamiltonians for
$A-a$ and $a$ nucleons, respectively. The same inter-nucleon interactions are consistently employed in each term of Eq.~(\ref{eq:formalism_40}).  Accordingly, the clusters' eigenstates $\ket{A-a \; \alpha_1 I_1^{\pi_1}T_1}$ and $\ket{a \; \alpha_2 I_2^{\pi_2}T_2}$
are obtained by NCSM diagonalization of their respective microscopic Hamiltonians $\hat{H}_{(A-a)}$ and $\hat{H}_{(a)}$. The same frequency and consistent model-space size are used in the HO expansions of the clusters and localized parts of the integration kernels. The size 
$N_{\rm max}$ of the HO model space is the same for states of the same parity, 
whereas it differs
by one unit for states of opposite parity.

While the NCSM/RGM formalism has been fully developed for single- ($a=1$)~\cite{Quaglioni:2009mn}, two- ($a=2$)~\cite{Navratil:2011ay} and three-nucleon ($a=3$) projectiles~\cite{Quaglioni:2012bm}, and can be also extended to $a=4$ projectiles as well as to three-body clusters~\cite{FB20proc}, in this work    
we limit ourselves to the 
$a=1$ case, where  
the inter-cluster antisymmetrizer is defined as
\begin{eqnarray}\label{eq:formalism_22}
  \hat{\mathcal{A}}_{\nu} & \equiv &
      \frac{1}{\sqrt{A}}\left( 1-\sum_{i=1}^{A-a}\hat{P}_{iA}\right),
\end{eqnarray} 
and $\hat{P}_{i,A}$ is the permutation operator exchanging the $i$-th particle 
in the target with the 
projectile nucleon, labeled by the index $A$.

\subsubsection{Orthogonalization in the NCSM/RGM}
\label{NCSMRGMkernels}
Here, we recall some of the details 
concerning the orthogonalization of the NCSM/RGM equations~(\ref{RGMeq}) 
that are useful for our further discussion of the NCSMC formalism. 

Because of the non-identical permutations in the inter-cluster antisymmetrizer, the channel states ${\mathcal A}_\nu\ket{\Phi_{\nu r}^{J^\pi T}}$ are not orthonormal to each other.
In general, we prefer to work with the orthonormalized binary-cluster states
\begin{equation}\label{eq:formalism_70}
  \sum_{\nu'}\int dr' {r'}^2 \; \mathcal{N}_{\nu \nu'}^{-\frac{1}{2}}(r,r')
                            \; \hat{\mathcal{A}}_{\nu'} \ket{\Phi_{\nu' r'}^{J^\pi T}} \; ,
\end{equation}
where we 
introduced the inverse square root of the NCSM/RGM norm kernel (\ref{N-kernel}). In the following we review how this as well as the square root of the norm kernel are obtained.
 
As anticipated in the previous section, the ``exchange'' term 
arising from the permutations in $\hat{\mathcal{A}}_\nu$
that differ from the identity 
are obtained using the HO expansion of Eq.~(\ref{eq:formalism_ff_82}). Hence, using Eqs.~(\ref{eq:formalism_22}) and (\ref{eq:formalism_84}), 
the $r$-space representation of the 
norm kernel 
can be written as
\begin{widetext}
\begin{align}
\label{eq:formalism_86}
\mathcal{N}^{J^\pi T}_{\nu \nu'}(r,r') 
& = \delta_{\nu\nu'}\frac{\delta(r-r')}{rr'} -(A-1)\sum_{n,n'}R_{n\ell}(r) \matrEL{\Phi_{\nu n}^{J^\pi T}}{\hat{P}_{A-1,A}}{\Phi_{\nu' n'}^{J^\pi T}} R_{n'\ell'}(r') \nonumber \\
&  = \delta_{\nu\nu'}\left[ 
            \frac{\delta(r-r')}{rr'} 
           -\sum_{nn' \in P} R_{n\ell}(r)\delta_{nn'}R_{n'\ell'}(r')
      \right]  +\sum_{nn' \in P} R_{n\ell}(r) \mathcal{N}^{J^\pi T}_{\nu n \nu' n'} R_{n'\ell'}(r') \; ,
\end{align}
\end{widetext}
%
where we introduced the model-space 
norm kernel:
\begin{align}
\label{eq:formalism_86.2}
 \mathcal{N}^{J^\pi T}_{\nu n \nu' n'} & =
      \delta_{\nu\nu'}\delta_{nn'} -(A-1)\sum_{n,n' \in P}R_{n\ell}(r) R_{n'\ell'}(r') \nonumber \\	
& \phantom{= \delta_{\nu\nu'}\delta_{nn'} -}
    \times \matrEL{\Phi_{\nu n}^{J^\pi T}}{\hat{P}_{A-1,A}}{\Phi_{\nu' n'}^{J^\pi T}} \;.
\end{align}
The last line of Eq.~(\ref{eq:formalism_86}) shows that the $r$-space representation
of the kernel is given by the convolution of the model-space kernel (second term) plus a correction
due to the finite size of the model space $P$ (first term).
Square and inverse-square roots $\mathcal{N}_{\nu \nu'}^{\pm \frac{1}{2}}(r,r')$ can then be defined in an analogous way as:
\begin{align}
&\mathcal{N}_{\nu \nu'}^{\pm \frac{1}{2}}(r,r')  \nonumber \\
& \quad = \delta_{\nu\nu'}\left[ 
                  \frac{\delta(r-r')}{rr'} 
                 -\sum_{nn' \in P} R_{n\ell}(r) \delta_{nn'}R_{n'\ell'}(r')
           \right] \nonumber \\
& \quad \phantom{=}     +\sum_{nn' \in P} R_{n\ell}(r) \mathcal{N}^{\pm \frac{1}{2}}_{\nu n \nu' n'} R_{n'\ell'}(r')\,,
\end{align}
where the model-space square and inverse square roots $\mathcal{N}^{\pm \frac12}_{\nu n \nu' n'}$ are obtained from the spectral theorem.

The NCSM/RGM 
Hamiltonian kernel 
within the orthonormal basis of Eq.~(\ref{eq:formalism_70}),
\begin{align}
\label{eq:formalism_90}
&\overline{\mathcal{H}}_{\nu \nu'}(r,r')  \\
&\quad =  \sum_{\mu\mu'}\int\!\! \int \!\! dy dy' {y}^2 {y'}^2 
                                    \mathcal{N}_{\nu \mu}^{-\frac{1}{2}}(r,y)
                                    \mathcal{H}_{\mu \mu'}(y,y')
                                    \mathcal{N}_{\mu' \nu'}^{-\frac{1}{2}}(y',r') \nonumber
\end{align}
is obtained from the hermitized Hamiltonian kernel,
\begin{align}
\label{eq:formalism_100}
\mathcal{H}_{\nu \nu'}(r,r')  & \!=\! 
	\matrEL{\Phi_{\nu r}^{J^\pi T  }}
           {\frac{1}{2}(
                        \hat{\mathcal{A}}^2 \hat {H} -  \hat {H} \hat{\mathcal{A}}^2
                       )
           }
           {\Phi_{\nu' r'}^{J^\pi T  }} \\
& \!=\!
 \matrEL{\Phi_{\nu r}^{J^\pi T}}
           {\hat {H} \! - \! \frac{1}{2}(
                                    \hat{H} \! \sum_i^{A-a} \hat{P}_{iA} 
                                       \!  - \! \sum_i^{A-a} \hat{P}_{iA} \hat{H}
                       )
           }
           {\Phi_{\nu' r'}^{J^\pi T}} ,  \nonumber
\end{align}
for which we have borrowed the same notation $\mathcal{H}_{\nu \nu'}(r,r')$ used previously in Eq.~(\ref{H-kernel}).

Finally, the orthogonalized RGM equations read 
\begin{equation}
{\sum_{\nu^\prime}\int dr^\prime r^{\prime\,2}} \overline{\mathcal{H}}_{\nu\nu^\prime\,}(r,r^\prime)\frac{\chi_{\nu^\prime} (r^\prime)}{r^\prime} = E\,\frac{\chi_{\nu} (r)}{r} \; , \label{RGMorteq} 
\end{equation}
with the wave functions of the relative motion $\chi_{\nu} (r)$ related to the original functions $\gamma_{\nu} (r)$ by
\begin{eqnarray}\label{eq:formalism_120}
  \frac{{\chi}_\nu(r)}{r}=\sum_{\nu'} \int dr' {r'}^2 
                              \mathcal{N}_{\nu \nu'}^{+\frac{1}{2}}(r,r') \frac{\gamma_{\nu'}(r')}{r'} \;.
\end{eqnarray}
For more details on the NCSM/RGM kernels we refer the interested reader to Ref.~\cite{Quaglioni:2009mn}.

\subsection{NCSMC}
\label{Subsec:NCSMC}
The NCSMC ansatz for the many-body wave function includes both $A$-body square-integrable and $(A-a,a)$ binary-cluster continuous basis states according to:
%
\begin{align}
\label{NCSMC_wav}
\ket{\Psi^{J^\pi T}_A} & =  \sum_\lambda c_\lambda \ket{A \lambda J^\pi T} \! +\! \sum_{\nu}\! \int \! dr \: r^2 
                               \frac{\gamma_{\nu}(r)}{r}
                               \hat{\mathcal{A}}_\nu\ket{\Phi_{\nu r}^{J^\pi T}}.
\end{align}
%
The resulting 
wave function~(\ref{NCSMC_wav}) is capable of describing efficiently both bound and unbound states. Indeed, the NCSM sector of the basis (eigenstates $\ket{A \lambda J^\pi T}$) provides an effective description of the short- to medium-range $A$-body structure, while the NCSM/RGM cluster states make the theory able to handle the scattering physics of the system. In other words, with the expansion~(\ref{NCSMC_wav}) one obtains the coupling of the NCSM with the continuum. Clearly, the NCSMC model space is overcomplete, but this is not a concern, as it will be shown in the following.  

\subsubsection{NCSMC equations}
The discrete $(c_\lambda)$ and 
continuous $(\gamma_{\nu} (r))$ unknowns of the NCSMC wave function are obtained as solutions of the coupled equations
\begin{eqnarray}\label{eq:formalism_110}
  \left(
     \begin{array}{cc}
        H_{NCSM} & \bar{h} \\
        \bar{h}  & \overline{\mathcal{H}} 
     \end{array}
  \right)
  \left(
     \begin{array}{c}
          c \\
		  {\chi}
     \end{array}
  \right)
  = 
  E
  \left(
     \begin{array}{cc}
        1 & \bar{g} \\
        \bar{g}  & 1 
     \end{array}
  \right)
  \left(
     \begin{array}{c}
          c \\ 
		  {\chi}
     \end{array}
  \right),
\end{eqnarray}
where $\chi_{\nu} (r)$ 
are the relative wave functions in the NCSM/RGM sector when working with the orthogonalized cluster channel states of
Eq.~(\ref{eq:formalism_70}). 
These are related to the original wave functions $\gamma_{\nu} (r)$ of Eq.~(\ref{NCSMC_wav}) by the relationship~(\ref{eq:formalism_120}). Note, however, that the $\chi_{\nu} (r)$ appearing in Eqs.~(\ref{RGMorteq}) and (\ref{eq:formalism_110}) are in general different, {\em i.e.}, they are solutions of different equations.

The NCSM sector of the Hamiltonian kernel is a diagonal matrix of the NCSM energy eigenvalues $E_{\lambda}$~(\ref{NCSM_eq}),
\begin{equation}\label{eq:formalism_130}
  (H_{NCSM})_{\lambda \lambda'}=
    \matrEL{A \lambda J^\pi T}{\hat{H}}{A \lambda' J^\pi T}=E_{\lambda}\delta_{\lambda \lambda'}\,,
\end{equation}
while $\overline{\mathcal{H}}$ is the orthogonalized NCSM/RGM kernel of Eq.~(\ref{eq:formalism_90}).
Because of the orthogonalization procedure of Sec.~\ref{NCSMRGMkernels}, 
both diagonal blocks in the NCSMC norm kernel $N$ are identities in their respective spaces
\begin{equation}\label{eq:formalism_140}
  N^{\lambda\lambda'}_{\nu r \nu' r'} = 
  \left(
     \begin{array}{cc}
        \delta_{\lambda\lambda'} & \bar{g}_{\lambda \nu'}(r') \\[2mm]
        \bar{g}_{\lambda' \nu}(r)  & \delta_{\nu\nu'}\frac{\delta(r-r')}{rr'} 
     \end{array}
  \right).
\end{equation}
The coupling between square-integrable and binary-cluster sectors of the model space is described by the cluster form factor 
\begin{equation}\label{eq:formalism_150}
  \bar{g}_{\lambda \nu}(r)=\sum_{\nu'}\int dr' {r'}^2 
                            \braket{A \lambda J^\pi T}
                                   {\hat{\mathcal{A}}_{\nu'} \Phi_{\nu' r'}^{J^\pi T  }}
                            \; \mathcal{N}_{\nu' \nu}^{-\frac{1}{2}}(r',r)
\end{equation}
in the norm kernel, and by the coupling form factor 
\begin{equation}\label{eq:formalism_160}
  \bar{h}_{\lambda \nu}(r)=\sum_{\nu'}\int dr' {r'}^2 
                            \matrEL{A \lambda J^\pi T}
                                   {\hat{H} \hat{\mathcal{A}}_{\nu'}}
                                   {\Phi_{\nu' r'}^{J^\pi T  }}
                            \; \mathcal{N}_{\nu' \nu}^{-\frac{1}{2}}(r',r).
\end{equation}
in the Hamiltonian kernel. Detailed expressions for these form factors 
are given in Appendix~\ref{Appendix_A}.
The calculation of $\braket{A \lambda J^\pi T}{\hat{\mathcal{A}}_{\nu} \Phi_{\nu r}^{J^\pi T  }}$ overlap matrix elements between NCSM wave functions and binary-cluster states was also discussed 
in Ref.~\cite{Navratil:2004tw}. We also note that by squaring the 
absolute value of these matrix elements and integrating over $r$, one obtains spectroscopic factors.

The NCSMC equations can be orthogonalized in an analogous way to that presented for the NCSM/RGM in Sec.~\ref{NCSMRGMkernels}. 
To define the square and inverse square root of the NCSMC norm in the $r$-space representation, we first rewrite 
Eq.~(\ref{eq:formalism_140}) as the convolution of the model-space norm kernel plus a correction for the
finite size of  the HO model-space $P$
\begin{eqnarray}\label{eq:formalism_170}
  \lefteqn{
             N^{\lambda\lambda'}_{\nu r \nu' r'} 
          } \nonumber \\[2mm]
  &=&  
  \left(
     \begin{array}{ccc}
        0 & &0 \\
        0  & & \delta_{\nu\nu'}\frac{\delta(r-r')}{rr'} - \delta_{\nu\nu'}R_{n\ell}(r) \delta_{nn'} R_{n'\ell'}(r')
     \end{array}
  \right) \nonumber \\[2mm]
  & & +
  \left(
     \begin{array}{cc}
        \delta_{\lambda \tilde\lambda} & 0 \\
        0  & R_{\nu r \tilde\nu n}
     \end{array}
  \right)
  N^{\tilde\lambda\tilde\lambda^\prime}_{\tilde\nu n\,\tilde\nu^\prime n^\prime } 
  \left(
     \begin{array}{cc}
        \delta_{\tilde\lambda^\prime \lambda^\prime} & 0 \\
        0  & R_{\nu^\prime r^\prime \tilde\nu^\prime n^\prime}
     \end{array}
  \right)\,,\nonumber\\
  &&
\end{eqnarray}
where the sum over the repeating indexes $\tilde\lambda, \tilde\nu, n, \tilde\lambda^\prime, \tilde\nu^\prime$, and $n^\prime$ is implied, the notation $R_{\nu r \tilde\nu n}$  stands for $R_{n\ell}(r)\delta_{\nu\tilde\nu}$, and the model-space NCSMC norm is given by:
\begin{align}
N^{\tilde\lambda\tilde\lambda^\prime}_{\tilde\nu n\,\tilde\nu^\prime n^\prime }  = 
\left(
     \begin{array}{cc}
        \delta_{\tilde\lambda \tilde\lambda^\prime} & \bar{g}_{\tilde\lambda \tilde\nu^\prime n^\prime} \\[2mm]
        \bar{g}_{\tilde\lambda^\prime \tilde\nu n} & \delta_{\tilde\nu \tilde\nu^\prime}\delta_{nn'}
     \end{array}
 \right)\,.
\end{align}
Here, the model-space cluster form factor is related to the $r$-space one through $\bar{g}_{\lambda \nu}(r)=\sum_n R_{nl}(r)\bar{g}_{\lambda \nu n}$ (as demonstrated in Appendix~\ref{Appendix_A}).  
Accordingly, the square and inverse square roots of $N$ can then be defined as:
\begin{eqnarray}\label{eq:formalism_180}
  \lefteqn{
             (N^{\pm \frac{1}{2}})^{\lambda\lambda'}_{\nu r \nu' r'}
          } \nonumber \\[2mm]
  & = & 
  \left(
     \begin{array}{ccc}
        0 &  &0 \\
        0  &  & \delta_{\nu\nu'}\frac{\delta(r-r')}{rr'} -  R_{n\ell}(r) \delta_{\nu\nu'}\delta_{nn'} R_{n'\ell'}(r')
     \end{array}
  \right) \nonumber \\[2mm]
   &  & + \left(
     \begin{array}{cc}
        \delta_{\lambda\tilde\lambda} & 0 \\
        0  & R_{\nu r \tilde\nu n}
     \end{array}
  \right) 
(N^{\pm\frac12})^{\tilde\lambda\tilde\lambda^\prime}_{\tilde\nu n\, \tilde\nu^\prime n^\prime}
 \left(
     \begin{array}{cc}
             \delta_{\tilde\lambda^\prime \lambda^\prime} & 0 \\
                     0  & R_{\nu^\prime r^\prime \tilde\nu^\prime n^\prime}
     \end{array}
  \right).
 \nonumber\\
  &&
\end{eqnarray}
Inserting the identity $N^{-\frac{1}{2}}N^{+\frac{1}{2}}$ in both left- and right-hand sides of Eq.~(\ref{eq:formalism_110}) and multiplying
by $N^{-\frac{1}{2}}$ from the left one obtains the orthogonalized NCSMC equations
\begin{equation}\label{eq:formalism_200}
  \overline{H}
  \left(
     \begin{array}{c}
          \bar{c} \\
		  \bar{\chi}
     \end{array}
  \right)
  = 
  E
  \left(
     \begin{array}{c}
          \bar{c} \\ 
		  \bar{\chi}
     \end{array}
  \right)\,,
\end{equation}
where the orthogonalized Hamiltonian is given by,
\begin{equation}\label{eq:formalism_210}
  \overline{H} =
  N^{-\frac{1}{2}} 
  \left(
     \begin{array}{cc}
        H_{NCSM} & \bar{h} \\
        \bar{h}  & \overline{\mathcal{H}}
     \end{array}
  \right)
  N^{-\frac{1}{2}},
\end{equation}
and the orthogonal wave functions by:
\begin{equation}\label{eq:formalism_220}
  \left(
     \begin{array}{c}
          \bar{c} \\
		  \bar{\chi}
     \end{array}
  \right)
  = 
  N^{+\frac{1}{2}} 
  \left(
     \begin{array}{c}
          c \\ 
		  {\chi}
     \end{array}
  \right).
\end{equation}
Finally, the ansatz~(\ref{NCSMC_wav}) in terms of the orthogonalized NCSMC wave function takes the form:
\begin{widetext}
\begin{eqnarray}\label{NCSMC_wave_orth}
\ket{\Psi^{J^\pi T}_A} 
&=& \sum_\lambda \ket{A \lambda J^\pi T} \left[\sum_{\lambda^\prime} (N^{-\frac{1}{2}})^{\lambda\lambda'} \bar{c}_{\lambda^\prime}
+\sum_{\nu^\prime} \! \int \! dr^\prime \: r^{\prime 2}  (N^{-\frac{1}{2}})^{\lambda}_{\nu^\prime r^\prime} \frac{\bar{\chi}_{\nu^\prime}(r^\prime)}{r^\prime}\right]
\nonumber \\
&+& \sum_{\nu\nu^\prime} \! \int \! dr \: r^2  \! \int \! dr^\prime \: r^{\prime 2} \hat{\mathcal{A}}_\nu\ket{\Phi_{\nu r}^{J^\pi T}}
\mathcal{N}_{\nu \nu'}^{-\frac{1}{2}}(r,r') 
\left[\sum_{\lambda^\prime} (N^{-\frac{1}{2}})^{\lambda'}_{\nu' r'}\bar{c}_{\lambda'} 
+ \sum_{\nu''}  \! \int \! dr'' \: r''^2  (N^{-\frac{1}{2}})_{\nu' r' \nu'' r''} \frac{\bar{\chi}_{\nu''}(r'')}{r''}\right] .
\end{eqnarray}
\end{widetext}

\subsubsection{Solving the NCSMC equations}\label{section_solving}

At large inter-cluster distances $r$, the clusters are assumed to interact 
through the Coulomb interaction only. Hence, the NCSMC equations are solved 
dividing the space into an internal region, $r \leqslant r_0$, and an external 
region, 
$r > r_0$, and applying the coupled-channel microscopic R-matrix method on 
a Lagrange mesh~\cite{R-matrix}. 
The separation radius $r = r_0$ must be large enough to ensure that the
wave function of the $A$-body states $\ket{A \lambda J^\pi T}$ vanishes when 
approaching the external region,
where the asymptotic behavior of the NCSMC solutions is described entirely by the radial wave functions
\begin{equation}\label{eq:formalism_230}
  u_\nu^{J^\pi T}(r)= C_\nu^{J^\pi T} W_\ell(\eta_\nu,\kappa_\nu r) ,
\end{equation}
and
\begin{align}
\label{eq:formalism_240}
u_\nu^{J^\pi T}(r) = \frac{i}{2}v_\nu^{-\frac{1}{2}}[\delta_{\nu i} H^-_\ell(\eta_\nu,\kappa_\nu r) 
                                               - S_{\nu i}^{J^\pi T} H^+_\ell(\eta_\nu,\kappa_\nu r)]
\end{align}
for bound and scattering states, respectively. 
Here, $W_l(\eta_\nu,\kappa_\nu r)$ are Whittaker functions and $H^\pm_l(\eta_\nu,\kappa_\nu r)$
are the incoming and outgoing Coulomb functions, with $v_\nu$ the speed, $\kappa_\nu$ the wave number, and $\eta_\nu$ the Sommerfeld parameter of the final state being studied. Asymptotic normalization constant for bound states and scattering matrix between initial $(i)$ and
final $(\nu)$ scattering states are denoted respectively with $C_\nu^{J^\pi T} $ and $S_{\nu i}^{J^\pi T}$.  
The functions $u_\nu^{J^\pi T}(r)$ stand for either the non-orthogonalized 
wave function $\chi_\nu(r)$ or for the orthogonalized $\bar{\chi}_\nu(r)$ according to which set of equations, Eq.~(\ref{eq:formalism_110}) or~(\ref{eq:formalism_200}), is being considered.   

One of the advantages of the microscopic R-matrix method is that the wave function 
in the internal region can be expanded on a set of 
square-integrable functions. 
A particularly convenient choice when dealing with non-local potentials, as in our case, is the set of Lagrange functions 
$f_{\nn}(r)$ associated with the shifted Legendre polynomials and defined 
on the mesh points $r_{\nn} \in (0,r_0)$~\cite{R-matrix}, labeled by the index $1\le\nn\le\NN$. 
When the Gauss quadrature approximation is adopted,
the Lagrange functions are orthogonal to each other. In addition, thanks to the Gauss quadrature approximation and the properties of the Lagrange functions, matrix elements of non-local potentials are proportional to the values of the non-local potentials at the mesh points.  
The number of mesh points $\NN$ has to be large enough to
guarantee an accurate representation of the wave functions in the internal region up to the matching radius $r_0$. Typically, 25 mesh points are sufficient to calculate a phase shift within six significant digits for $r_0=15$ fm.

The matching between 
internal and 
external regions, and hence 
the imposition of the asymptotic behavior of 
Eqs.~(\ref{eq:formalism_230}) and/or (\ref{eq:formalism_240}), 
is ensured by 
the Bloch surface operator (here generalized to account for the $A$-body square-integrable sector of our basis)
\begin{equation}\label{eq:formalism_260}
  \hat{L}_\nu =
  \left(
     \begin{array}{ccc}
        0 &  &0 \\
        0  &  & \frac{\hbar^2}{2\mu_\nu} \delta(r-a) (\frac{d}{dr} - \frac{B_\nu}{r})
     \end{array}
  \right)
\end{equation}
and solving the Bloch-Schr\"odinger equations
\begin{equation}\label{eq:formalism_270}
  (\overline{H}+\hat{L}-E)
  \left(
     \begin{array}{c}
          \bar{c} \\
		  \bar{\chi}
     \end{array}
  \right)
  = 
  \hat{L}
  \left(
     \begin{array}{c}
          \bar{c} \\ 
		  \bar{\chi}
     \end{array}
  \right).
\end{equation}
The operator $\overline{H}+\hat{L}$ is Hermitian when the boundary parameter $B_\nu$ is real.
Because of the Bloch operator, the wave function in the right-hand side
of Eq.~(\ref{eq:formalism_270}) can be replaced by its asymptotic behavior.
When searching for  
bound states, $B_\nu$ is chosen in such a way that the 
right-hand side vanishes, and one is left with the diagonalization problem:
\begin{equation}\label{eq:formalism_280}
  (\overline{H}+\hat{L})
  \left(
     \begin{array}{c}
          \bar{c} \\
		  \bar{\chi}
     \end{array}
  \right)
  = 
  E
  \left(
     \begin{array}{c}
          \bar{c} \\ 
		  \bar{\chi}
     \end{array}
  \right).
\end{equation}
For 
scattering states, 
the scattering matrix and the scattering wave functions are
computed by solving Eq.~(\ref{eq:formalism_270}) with the boundary parameter $B_\nu=0$
for each value of the relative kinetic energy $E_{kin}$ of the projectile-target system. 
The phase shifts $\delta(E_{kin})$ can then be extracted from the S-matrix. 
Energetically open and closed channels are treated on equal footing. 

\section{Application to $\elemA{7}{He}$}
\label{Sec:7He}

The $\elemA{7}{He}$ nucleus is a particle-unstable system with a $J^\pi T=3/2^-\; 3/2$ ground state
lying at $0.430(3) \mev$ \cite{Stokes_1967,Cao2012} above the $\elemA{6}{He} + n$ threshold 
and an excited $5/2^-$ resonance centered at $3.35 \mev$, 
which mainly decays to $\alpha+3n$ (as discovered in the 
pioneering work of Ref.~\cite{Korsheninnikov_1998}).
While there is a general consensus on the $5/2^-$ state,
discussions are still open for the other excited states.
In particular, the existence of a low-lying  ($E_R{\sim} 1$ MeV) narrow ($\Gamma \leq 1$ MeV)  $1/2^-$ state 
has been advocated by many
experiments \cite{Markenroth_2001,Meister_2002,Skaza_2006,Ryezayeva_2006,Lapoux_2006}
(most of them using knockout reactions with a $\elemA{8}{He}$ beam on a carbon target), 
while it was not confirmed in several others
\cite{Bohlen_2001,Rogachev_2004,Wuosmaa_2005,Wuosmaa_2008,Denby_2008,Aksuytina_2009}.
This contradictory situation arises from the main experimental difficulty 
of measuring the properties of $\elemA{7}{He}$ excited states
 in the presence of a three-body background of $^6{\rm He}$ plus $n$
(coming from the particle decay of $\elemA{7}{He}$) plus a third outgoing particle
involved in the reaction used to produce $\elemA{7}{He}$.
In addition, as it has been pointed out in one of the most recent experimental works 
\cite{Aksuytina_2009}, some of the earlier data could have been affected 
by background noise coming from the interaction with the carbon target,
while a polypropylene $(CH_2)_n$ target would reduce the background contamination.
The presence of a low-lying $1/2^-$ state has also been excluded at the $90\%$ 
confidence level by a study on the isobaric analog states of $\elemA{7}{He}$
in $\elemA{7}{Li}$~\cite{Boutachkov_2005}. According to this work, a broad 
$1/2^-$ resonance at ${\sim}3.5 \mev$ with a width $\Gamma{\sim}10 \mev$ 
fits the data the best.
Neutron pick-up and proton-removal reactions~\cite{Wuosmaa_2005, Wuosmaa_2008} 
suggest instead a $1/2^-$ resonance
at about $3 \mev$ with a width $\Gamma \approx 2 \mev$.

From a theoretical point of view, $^7$He is an ideal system to showcase new achievements made possible by a unified {\em ab initio} approach to nuclear bound and continuum states such as the NCSMC. Since $^7$He is unbound, it cannot be reasonably described within the NCSM. One could calculate its properties using the NCSM/RGM within an $^6$He+$n$ binary-cluster expansion. However, the $^6$He nucleus is weakly bound 
and all its excited states are unbound. Consequently, a limitation to just a few lowest $^6$He eigenstates in the NCSM/RGM expansion would be questionable especially because, except for the lowest $2^+$ state, all 
other $^6$He excited states are either broad resonances or simply states in the continuum. As we will show in the following, with the NCSMC these challenges are overcome. Finally, for this study we use the SRG evolved~\cite{SRG,Roth_SRG,Roth_2010,Bogner_2010} chiral N3LO $NN$ potential ($500 \mev$ cutoff) of Refs.~\cite{Entem:2003ft,Machleidt:2011zz}.  For the time being, the induced and initial chiral three-nucleon interactions are not included in the calculations, therefore our results depend on the low-momentum SRG evolution parameter $\Lambda$. However, 
by selecting $\Lambda = 2.02$ fm$^{-1}$, we obtain very realistic binding energies for the lightest nuclei, e.g., the $^4$He (see Table~\ref{tab:NCSM_He_gs}) and, more importantly for the present investigation, the $^6$He.  Consequently, this choice of $NN$ potential allows us to perform qualitatively and quantitatively meaningful calculations for $^7$He that can be compared to experiment.
In the following sections, we discuss the convergence of the NCSMC calculation and compare it to the corresponding NCSM and 
NCSM/RGM results. We will also address the controversial issue of a low-lying $1/2^-$ resonance in $^7$He.

\subsection{$^6$He and $^7$He NCSM calculations}
\label{subsec:He_NCSM}

\begin{figure}[!ht]
\begin{center}
\includegraphics[clip=,width=0.45\textwidth]{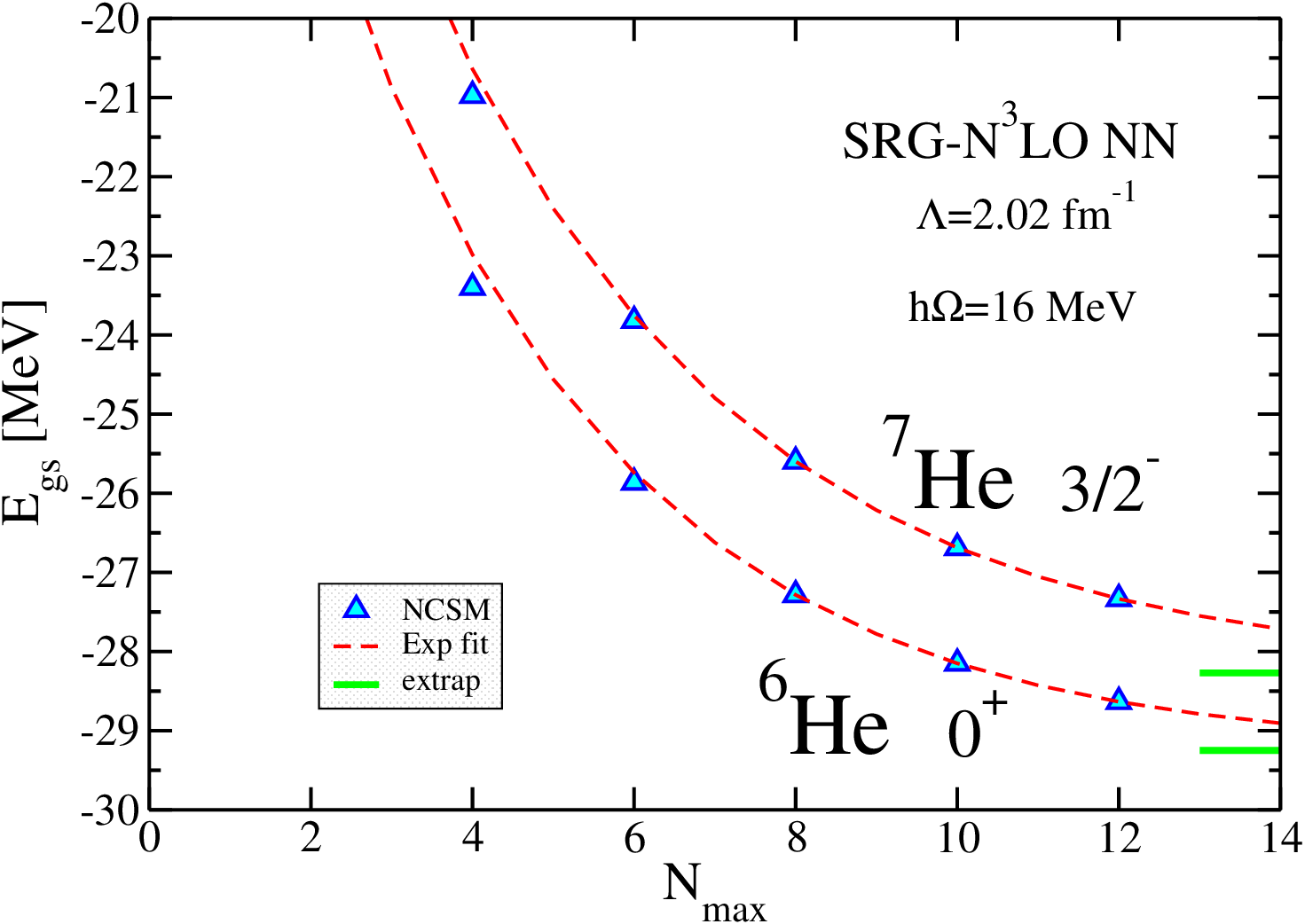}
\end{center}
\caption{(color online). Basis size $N_{max}$ dependence of $^6$He and $^7$He ground-state energies calculated within the NCSM. The SRG-N$^3$LO $NN$ potential with $\Lambda=2.02$ fm$^{-1}$ and the HO frequency of $\hbar\Omega{=}16$ MeV were used. Exponential extrapolations from the last three $N_{\rm max}$ points is shown.}
\label{7He_6He_gs_NCSM}
\end{figure}

We begin our discussion of results with NCSM calculations for $^6$He that will generate eigenstates needed as input for the subsequent NCSM/RGM and NCSMC investigations of $^7$He. 
\begin{table}[t]
\begin{center}
\begin{ruledtabular}
\begin{tabular}{c|ccc}
 $E_{\rm g.s.}$ [MeV]           & $^4$He   &  $^6$He     &  $^7$He      \\
\hline
NCSM $N_{\rm max}{=}12$ &  -28.05   &  -28.63      &  -27.33       \\
NCSM extrap.                   & -28.22(1)& -29.25(15) &  -28.27(25) \\  
Expt.                                &  -28.30   & -29.27       &  -28.84       \\
\end{tabular} 
\end{ruledtabular}
\caption{Ground-state energies of $^{4,6,7}$He in MeV. NCSM calculations were performed with the SRG-N$^3$LO $NN$ potential with $\Lambda=2.02$ fm$^{-1}$. The HO frequency $\hbar\Omega{=}16$ MeV was used in the shown $N_{\rm max}{=}12$ calculations. Exponential extrapolation was employed.}
\label{tab:NCSM_He_gs}
\end{center}
\end{table}
\begin{figure}[b]
\begin{center}
\includegraphics[clip=,width=0.45\textwidth]{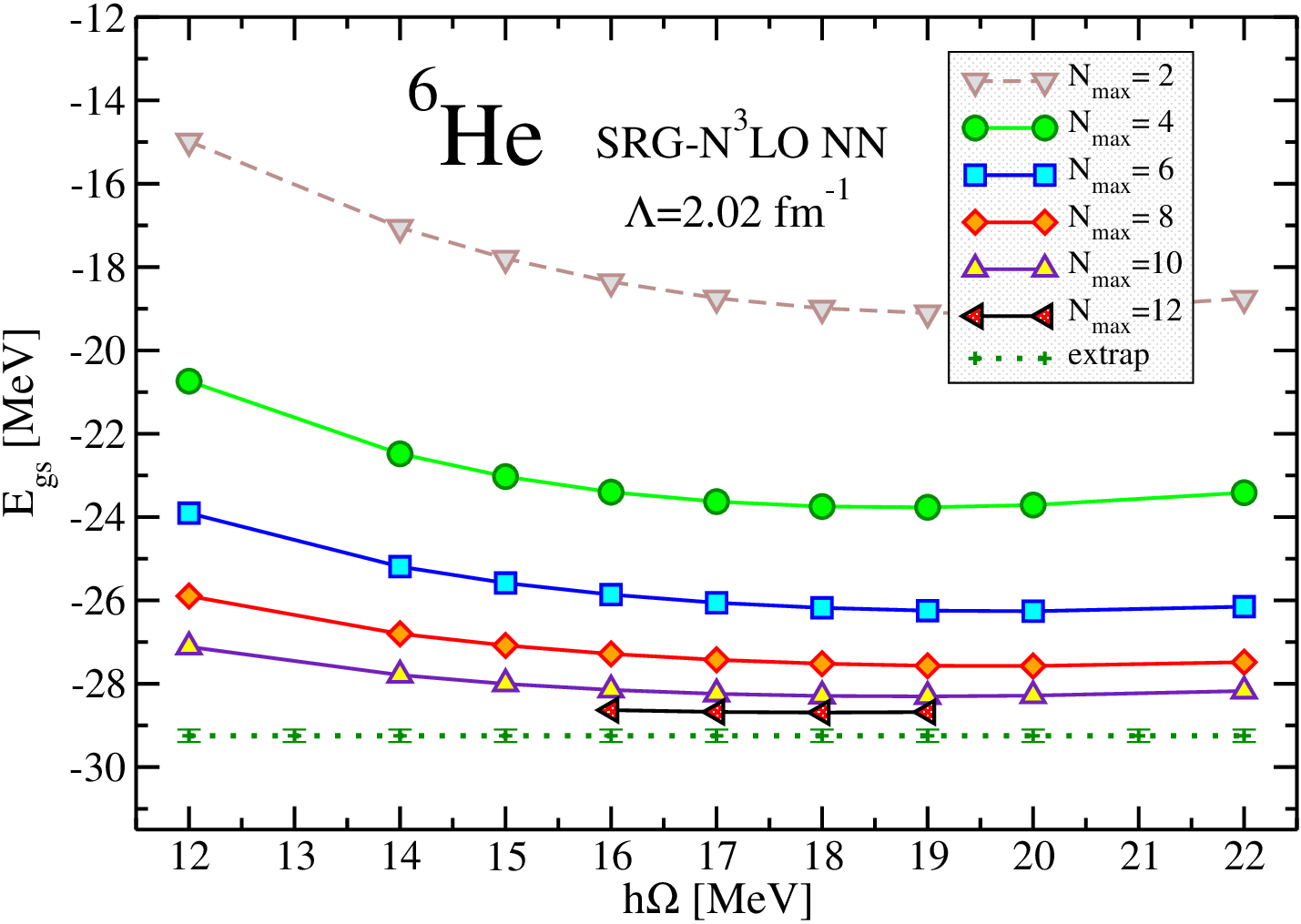}
\end{center}
\caption{(color online). Ground-state energy of $^{6}$He calculated within the NCSM using the SRG-N$^3$LO $NN$ potential with $\Lambda=2.02$ fm$^{-1}$. The dependence on the HO frequency for different $N_{\rm max}$ basis sizes is shown. The points with error bars represent the results of the exponential extrapolation.}
\label{6He_gs}
\end{figure}
Our calculated $^6$He ground-state energies for a range of HO frequencies and various basis sizes ($N_{\rm max}$ values) are presented in Fig.~\ref{6He_gs}. The variational NCSM calculations converge rapidly and can be easily extrapolated to $N_{\rm max}\rightarrow\infty$ using, e.g., an exponential function of the type $E(N_{\rm max})=E_\infty+a\; e^{-b N_{\rm max}}$. 
Results of such an extrapolation 
are shown in Fig.~\ref{7He_6He_gs_NCSM} where the $N_{\rm max}{=}8,10$ and $12$ points were used to determine the fitting parameters. The extrapolated ground-state energy with its error estimate, based on extrapolations at other frequencies and different point selections, and the calculated energy at $N_{\rm max}=12$, $\hbar\Omega=16$ MeV are given in Table~\ref{tab:NCSM_He_gs}. The calculated $^6$He g.s.\ energy agrees quite well with experiment
on both the absolute value and the separation with respect to the $^4$He$+2n$ threshold. 

\begin{figure}[!ht]
\begin{center}
\includegraphics[clip=,width=0.45\textwidth]{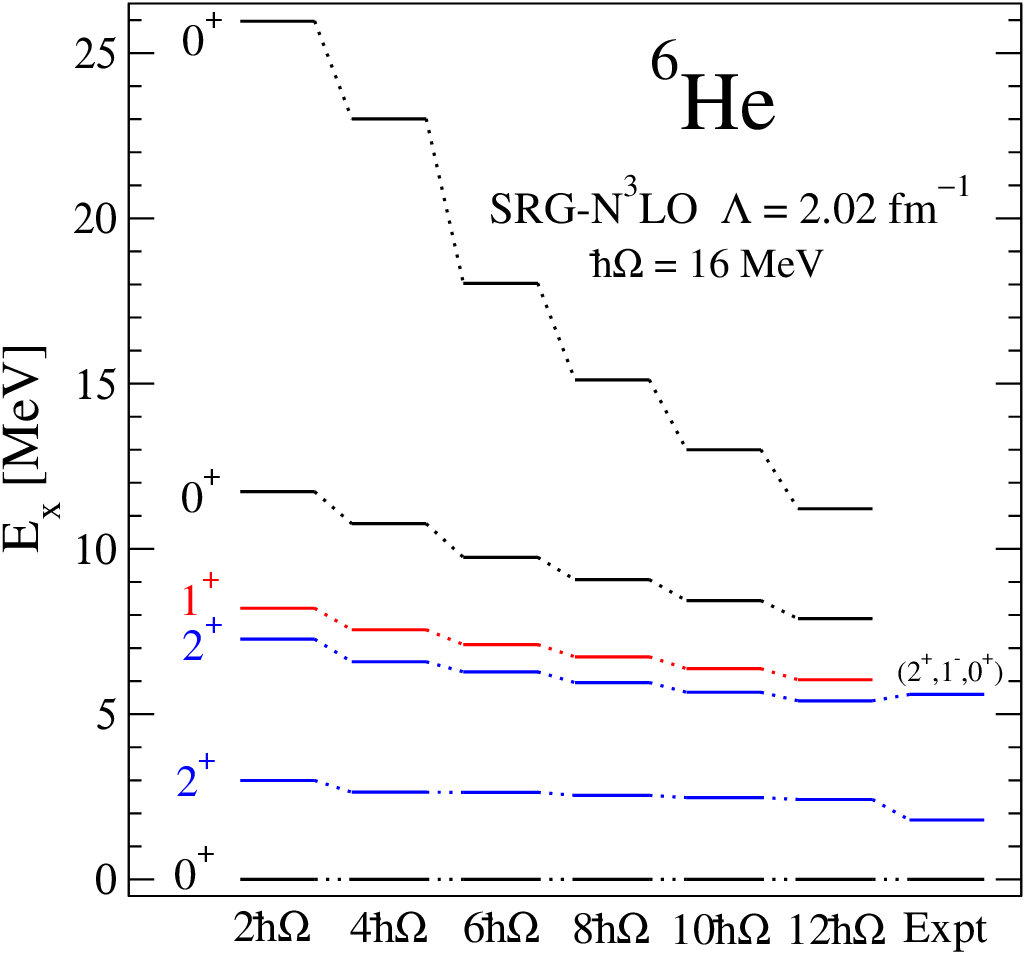}
\end{center}
\caption{(color online). Dependence of $^6$He excitation energies on the size of the basis $N_{max}$. NCSM calculations were performed with the SRG-N$^3$LO $NN$ potential with $\Lambda=2.02$ fm$^{-1}$ and the HO frequency of $\hbar\Omega{=}16$ MeV.}
\label{6He_exct}
\end{figure}

As shown in Fig.~\ref{6He_gs}, at $N_{\rm max}=12$ the dependence of the $^6$He g.s.\ energy on the HO frequency is flat in the range of $\hbar\Omega\sim 16 - 19$ MeV. In general, when working within an HO basis, lower frequencies are better suited for the description of unbound systems. Therefore, we choose $\hbar\Omega=16$ MeV for the calculation of the other $^6$He eigenstates that will be used as input for the NCSM/RGM and 
NCSMC investigations of the $^7$He nucleus. 
At the same time, we also performed 
NCSM/RGM and NCSMC calculations 
with $\hbar\Omega=19$ MeV,  to 
test the stability of our results against this parameter. Calculated $^6$He excitation energies for basis sizes up to $N_{\rm max}=12$ are shown in Fig.~\ref{6He_exct}.  The $^6$He nucleus is a is a weakly-bound Borromean system. All its excited states are unbound, and, except for the lowest $2^+$, either broad resonances or states in the continuum. The excitation energy of the $2^+_1$ state is fairly stable with respect to the basis size of our NCSM calculations. The higher excited states, however, drop in energy with increasing $N_{\rm max}$ with the most dramatic example being the multi-particle-hole $0^+_3$ state. This spells a potential difficulty for the NCSM/RGM calculations as, with increasing density of $^6$He states at low energies, a truncation of the model space to include just the few lowest eigenstates becomes questionable. In addition, in a NCSM/RGM study of $^7$He one should also consider the contribution of binary-cluster states in which the neutron is coupled to negative-parity states of $^6$He, where similar issues arise. 

Next we performed NCSM calculations for $^7$He ground 
and excited states, which will serve as input to the NCSMC calculations described in the next section. The calculated g.s.\ energy for different basis sizes is shown in Fig.~\ref{7He_6He_gs_NCSM} together with the exponential extrapolation and the $^6$He g.s.\ energies discussed earlier. The largest-space values and the extrapolated energies are also given in Table~\ref{tab:NCSM_He_gs}. The NCSM calculation predicts $^7$He unbound in agreement with experiment. However, the resonance energy with respect to the $^6{\rm He}+n$ threshold appears overestimated (contrary to the $^6$He$\leftrightarrow ^4$He$+2n$ case). Obviously, it is not clear that the ad hoc exponential extrapolation is valid for the unbound states. Or, it may have a sizable systematic uncertainty compared to the bound-state case. 
Nevertheless, 
the differences between 
$N_{\rm max}=12$ and 
extrapolated energies 
suggest that 
the fastest convergence rate is obtained for the strongly-bound $^4$He and the slowest for the unbound $^7$He, as one would expect. Finally, we note that no information on the width of the resonance can be obtained from the NCSM calculation, which is performed in a square-integrable HO basis. 

\begin{table}[!ht]
\begin{center}
\begin{ruledtabular}
\begin{tabular}{cc|c|c|cc|c}
$^7$He $J^\pi$  & $^6$He${-}n(lj)$      & NCSM  & CK      & VMC & GFMC & Exp. \\
\hline
$3/2^-_1$         &  $0^+{-}p\frac{3}{2}$ & 0.56     & 0.59   & 0.53   & 0.565 & 0.512(18)\cite{Cao2012} \\
                         &                                   &             &           &           &           & 0.64(9) \cite{Beck2007} \\
                         &                                   &             &           &           &           & 0.37(7) \cite{Wuosmaa_2005} \\
$3/2^-_1$         &$2^+_1{-}p\frac{1}{2}$& 0.001    &  0.06  &  0.006&          &   \\
$3/2^-_1$         &$2^+_1{-}p\frac{3}{2}$& 1.97     &  1.15  &  2.02   &          &   \\
$3/2^-_1$         &$2^+_2{-}p\frac{1}{2}$&  0.12    &        &  0.09      &           &   \\
$3/2^-_1$         &$2^+_2{-}p\frac{3}{2}$&  0.42    &        &  0.30      &           &   \\
$1/2^-$            & $0^+{-}p\frac{1}{2}$&  0.94    & 0.69   &  0.91     &           &   \\
$1/2^-$            &$2^+_1{-}p\frac{3}{2}$&  0.34   & 0.60   &  0.26    &           &   \\
$1/2^-$            &$2^+_2{-}p\frac{3}{2}$&   0.93   &           &            &           &   \\
$5/2^-$            &$2^+_1{-}p\frac{1}{2}$&   0.77   & 0.85   &  0.81    &          &   \\
$5/2^-$            &$2^+_1{-}p\frac{3}{2}$&   0.49   & 0.52   &  0.37    &           &   \\
$5/2^-$            &$2^+_2{-}p\frac{1}{2}$&   0.26   &           &            &           &   \\
$5/2^-$            &$2^+_2{-}p\frac{3}{2}$&   1.30   &           &            &           &   \\
$3/2^-_2$         &   $0^+{-}p\frac{3}{2}$&  0.06    &  0.06   &  0.05  &          &   \\
$3/2^-_2$         &$2^+_1{-}p\frac{1}{2}$&  1.10     &  1.05   &  1.07  &         &   \\
$3/2^-_2$         &$2^+_1{-}p\frac{3}{2}$&   0.08   &   0.32   &  0.03   &           &   \\
$3/2^-_2$         &$2^+_2{-}p\frac{1}{2}$&   0.03    &           &            &           &   \\
$3/2^-_2$         &$2^+_2{-}p\frac{3}{2}$&   0.25    &           &            &           &   \\
\end{tabular}
\end{ruledtabular}
\caption{NCSM spectroscopic factors compared to Cohen-Kurath (CK)~\cite{CK} 
and VMC/GFMC~\cite{GFMC,Brida2011,Wiringa} calculations and experiment. NCSM 
calculations were performed with the SRG-N$^3$LO $NN$ potential with 
$\Lambda{=}2.02$ fm$^{-1}$, $N_{\rm max}{=}12$ and the HO frequency of 
$\hbar\Omega{=}16$ MeV. The CK results should in principle be still multiplied by $A/(A{-}1)$ to correct for the center of mass motion.}
\label{tab:specfac}
\end{center}
\end{table}
However, we can study the structure of the $^7$He NCSM eigenstates by evaluating their overlap functions with $^6$He$+n$ binary-cluster channels. These overlap functions, or cluster form factors, $g_{\lambda\nu}(r)$ [see Eqs.~(\ref{eq:formalism_150}), (\ref{eq:formalism_ff_2}), (\ref{eq:formalism_ff_2.2})] are also one of the inputs to the NCSMC calculations. By integrating $g_{\rm \lambda\nu}^2(r)$ over $r$, we obtain the spectroscopic factors summarized in Table~\ref{tab:specfac}. Note that there we use an alternative coupling scheme [compared to Eq.~(\ref{eq:formalism_20})] more commonly used in the literature for spectroscopic factors. Overall, we find a very good agreement with the variational Monte Carlo (VMC) and GFMC results as well as with the latest experimental value for the ground state~\cite{Cao2012}. Interesting features to notice are the spread of the $3/2^-$ g.s.\ wave function over all three considered $^6$He states with a dominance of the $2^+_1$ and the about equal spread of $1/2^-$ $^7$He excited state between the $^6$He $0^+$ and $2^+_2$ states. The $^7$He $5/2^-$ state has about the same contributions from the $2^+_1$ and $2^+_2$ $^6$He states with the former of an almost pure $s{=}5/2$ component (with $s$ the channel spin defined in Eq.~(\ref{eq:formalism_20})). Though spectroscopic factors are not observable, they provide valuable information on the structure of the wave function. In the present study, overlap functions and spectroscopic factors are not the final products to be compared to experiment, but rather inputs to more sophisticated NCSMC calculations.     

\subsection{$^7$He  NCSM/RGM and NCSMC calculations}
\label{subsec:7He_NCSMC}

In the following, we present NCSMC calculation for the $^7$He nucleus performed within a model space containing the six lowest negative-parity ($3/2^-_1,1/2^-,5/2^-,3/2^-_2,3/2^-_3,3/2^-_4$) and four lowest positive-parity ($1/2^+,5/2^+_1,3/2^+,5/2^+_2$) NCSM eigenstates of $^7$He plus $n+^6$He NCSM/RGM binary-cluster channels including up to the three lowest eigenstates of $^6$He, {\em i.e.}\ $0^+, 2^+_1$, and  $2^+_2$. For the sake of comparison, we will also present results obtained by retaining only the binary-cluster portion of such a model space [{\em i.e.}, only the second term in Eq.~(\ref{NCSMC_wav}) or, equivalently, the ansatz~(\ref{eq:formalism_10})] and solving the orthogonalized NCSM/RGM equations of Eq.~(\ref{RGMorteq}).

%
\begin{figure}[!ht]
\begin{center}
\includegraphics[clip=,width=0.45\textwidth]{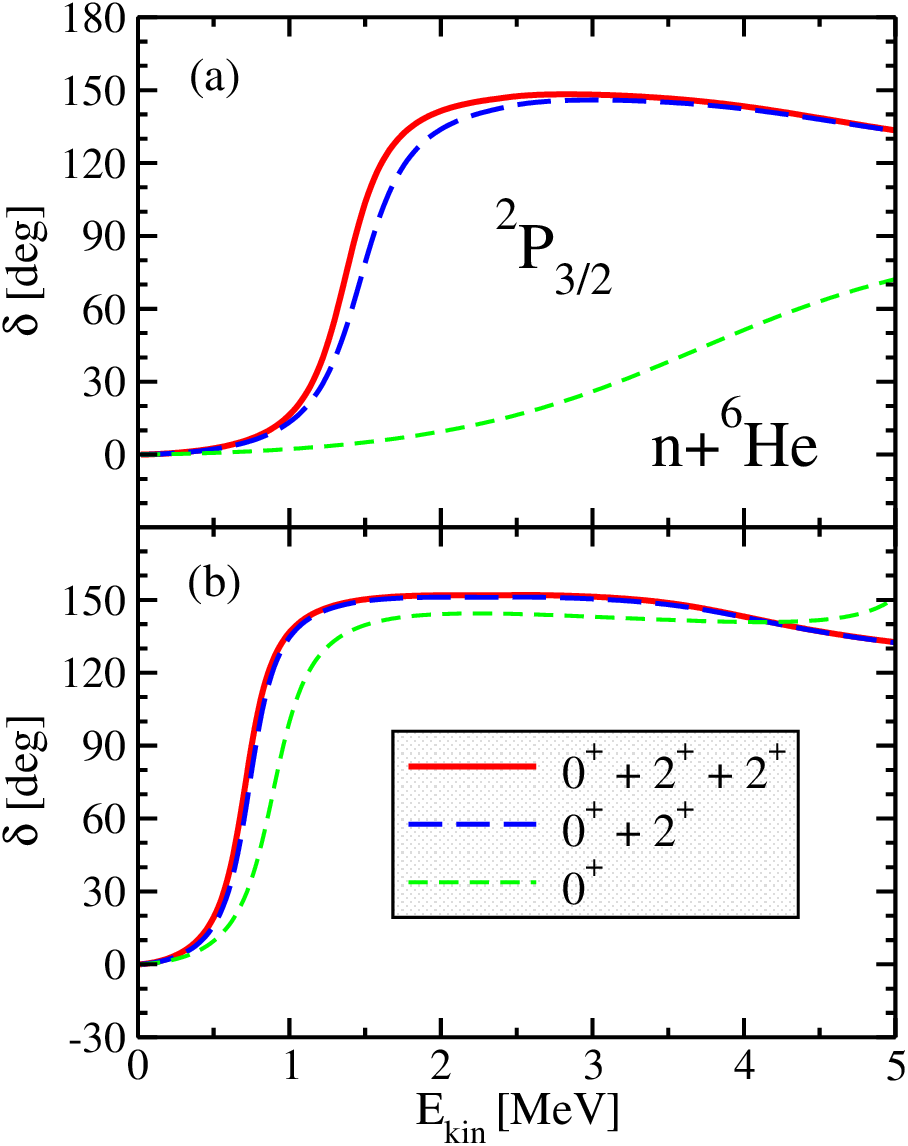}
\end{center}
\caption{(color online). Dependence of the NCSM/RGM (a) and NCSMC (b) $\elemA{6}{He} + n$ phase shifts of the $^7$He $3/2^-$ ground state
on the number of $^6$He states included in the binary-cluster basis. The short-dashed
green curve, the dashed blue curve and the solid red curve correspond to calculations
with $\elemA{6}{He}$ $0^+$ ground state only, $0^+,2^+$ states and $0^+,2^+,2^+$ states, respectively. The SRG-N$^3$LO $NN$ potential with $\Lambda=2.02$ fm$^{-1}$, the $N_{\rm max}{=}12$ basis size and the HO frequency of $\hbar\Omega{=}16$ MeV were used. See text for further details.}
\label{6He_n_target_states}
\end{figure}
We start by studying the dependence of the $3/2^-$ g.s.\ phase shifts on the number of $^6$He eigenstates included in the
NCSM/RGM [panel (a)] and NCSMC [panel (b)] calculations, shown in Fig.~\ref{6He_n_target_states}. 
Here, the channels are denoted using the standard notation ${}^{2s+1}\ell_{J}$, {\em e.g.}, $^2P_{3/2}$ for the g.s.\ resonance, 
with the quantum numbers $s, \ell$ and $J$ defined as in Sec.~\ref{Subsec:NCSMRGM}, Eq.~(\ref{eq:formalism_20}).  We observe that the NCSM/RGM calculation with the $^6$He target restricted to its ground state does not produce a $^7$He $3/2^-$ resonance (the phase shift does not reach 90 degrees and is less than 70 degrees up to 5 MeV). A $^2P_{3/2}$ resonance does appear once $n+^6$He$(2^+_1)$ channel states are coupled to the basis, and the 
resonance position further moves to lower energy with the inclusion of the second $2^+$ state of $^6$He. 
On the contrary, the NCSMC calculation with only the ground state of $^6$He already produces the $^2P_{3/2}$ resonance.  
In fact, this NCSMC model space is sufficient to 
obtain the $\elemA{7}{He}$ $3/2^-$ g.s.\ resonance at about $1 \mev$ above threshold, which is lower than the NCSM/RGM prediction of $1.39 \mev$ when three $\elemA{6}{He}$ states are included. Adding the first $2^+$ state of $^6$He generates a modest shift of the resonance to a still lower energy while the $2^+_2$ state of $^6$He 
has no significant influence [see Fig.~\ref{6He_n_target_states}, panel (b)]. We further observe that the difference of about 0.7 MeV between the NCSM/RGM and NCSMC results for the resonance position 
is due to additional correlations in the wave function brought about by the $^7$He eigenstates that are coupled to the neutron-$^6$He binary-cluster states in the NCSMC.  Indeed, such $A=7$ eigenstates (in the present calculation four $3/2^-$ states, of which only the $3/2^-_1$ 
produces a substantial effect on the $^2P_{3/2}$ resonance) 
have the practical effect of compensating for higher excited states of the $^6$He target omitted in the NCSM/RGM sector of the basis. These omitted $^6$He states 
include both positive-parity, some of which are shown in Fig.~\ref{6He_exct}, and 
negative-parity excitations such as, {\em e.g.}, the $1^-$ soft dipole excitation {\em etc.} While NCSM/RGM calculations with a large number of excited states of the target or projectile can become prohibitively expensive, the coupling of a few square-integrable NCSM eigenstates of the composite system is straightforward. Because of this, the NCSMC approach offers a superior rate of convergence and is 
much more efficient, as demonstrated in Fig.~\ref{6He_n_target_states}. 

\begin{figure}[t]
\begin{center}
\includegraphics[clip=,width=0.45\textwidth]{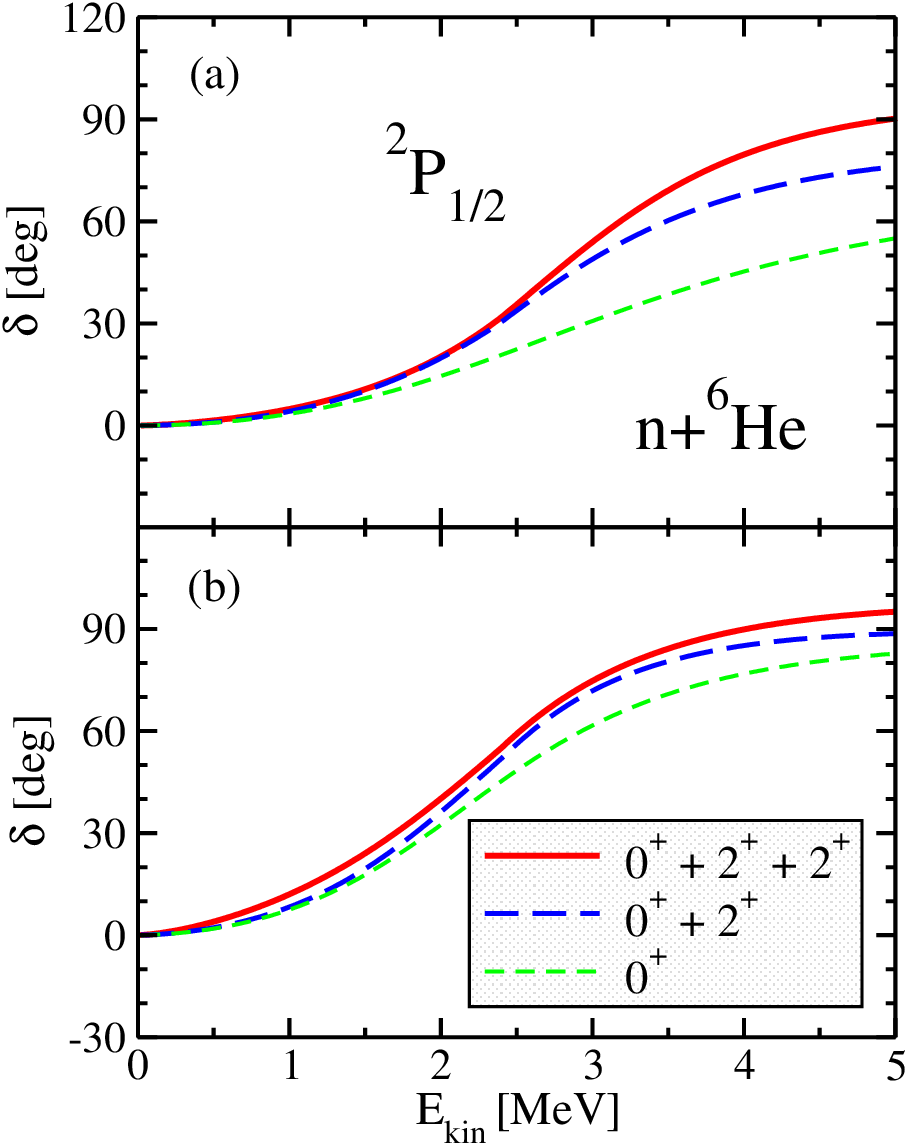}
\end{center}
\caption{(color online). Dependence of the NCSM/RGM (a) and NCSMC (b) $\elemA{6}{He} + n$ phase shifts of the $^7$He $1/2^-$ excited state
on the number of $^6$He states included in the binary-cluster basis. The short-dashed
green curve, the dashed blue curve and the solid red curve correspond to calculations
with $\elemA{6}{He}$ $0^+$ ground state only, $0^+,2^+$ states and $0^+,2^+,2^+$ states, respectively. See Fig.~\ref{6He_n_target_states} for further details.}
\label{6He_n_target_states_1m}
\end{figure}
A similar, although less dramatic, difference between 
NCSM/RGM and 
NCSMC calculations is shown in Fig.~\ref{6He_n_target_states_1m} for the $1/2^-$ excited state of $^7$He. Here, the $^2P_{1/2}$ resonance is quite broad with a slowly increasing phase shift. It is interesting to note that the $1/2^-$ state couples strongly to the $2^+_2$ state of $^6$He (the spectroscopic factor is large, see Table~\ref{tab:specfac}). This causes 
a small but visible shift of the $^2P_{1/2}$ phase shift when this state is added 
to the NCSMC calculation in panel (b)  (full vs.\ dashed line). The $1/2^-$ state 
presents a significant overlap also with the $0^+$ and $2^+_1$ states of $^6$He, 
and this is the reason 
of its broadness.     

\begin{figure}[t]
\begin{center}
\includegraphics[clip=,width=0.45\textwidth]{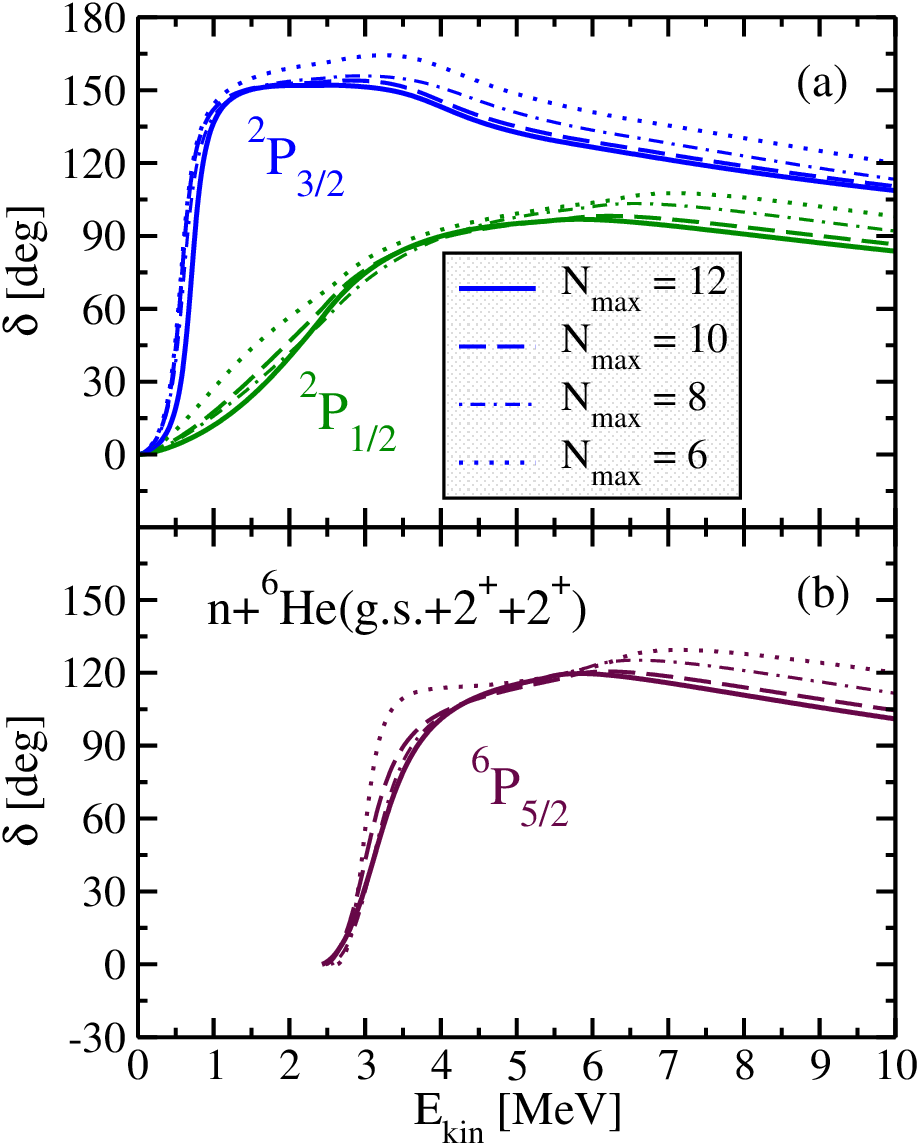}
\end{center}
\caption{(color online). Dependence of the NCSMC $\elemA{6}{He} + n$ phase shifts of $^7$He $3/2^-$ (a), $1/2^-$ (a) and $5/2^-$ (b) states
on the size of the HO expansion $N_{max}$. The $\elemA{6}{He}$ $0^+,2^+,2^+$ states were included in the binary-cluster basis. The SRG-N$^3$LO $NN$ potential with $\Lambda=2.02$ fm$^{-1}$ and the HO frequency of $\hbar\Omega{=}16$ MeV were used.}
\label{6He_n_nmax_dep}
\end{figure}
In Fig.~\ref{6He_n_nmax_dep}, we present the dependence of the NCSMC 
$^2P_{3/2}$, $^2P_{1/2}$ [panel (a)] and $^6P_{5/2}$ [panel (b)] phase shifts on the size of the HO 
basis in the range $6\le N_{\rm max} \le 12$. 
While the variation between 
$N_{\rm max} = 6$ and $N_{\rm max}=8$ curves is substantial, it becomes quite small between $N_{\rm max}=10$ and $N_{\rm max}=12$ results. Based on this, we do not expect 
that an $N_{\rm max}=14$ calculation,  which at this time is computationally out of reach, 
would significantly change the present $N_{\rm max}=12$ picture. 

\begin{figure}[!ht]
\begin{center}
\includegraphics[clip=,width=0.45\textwidth]{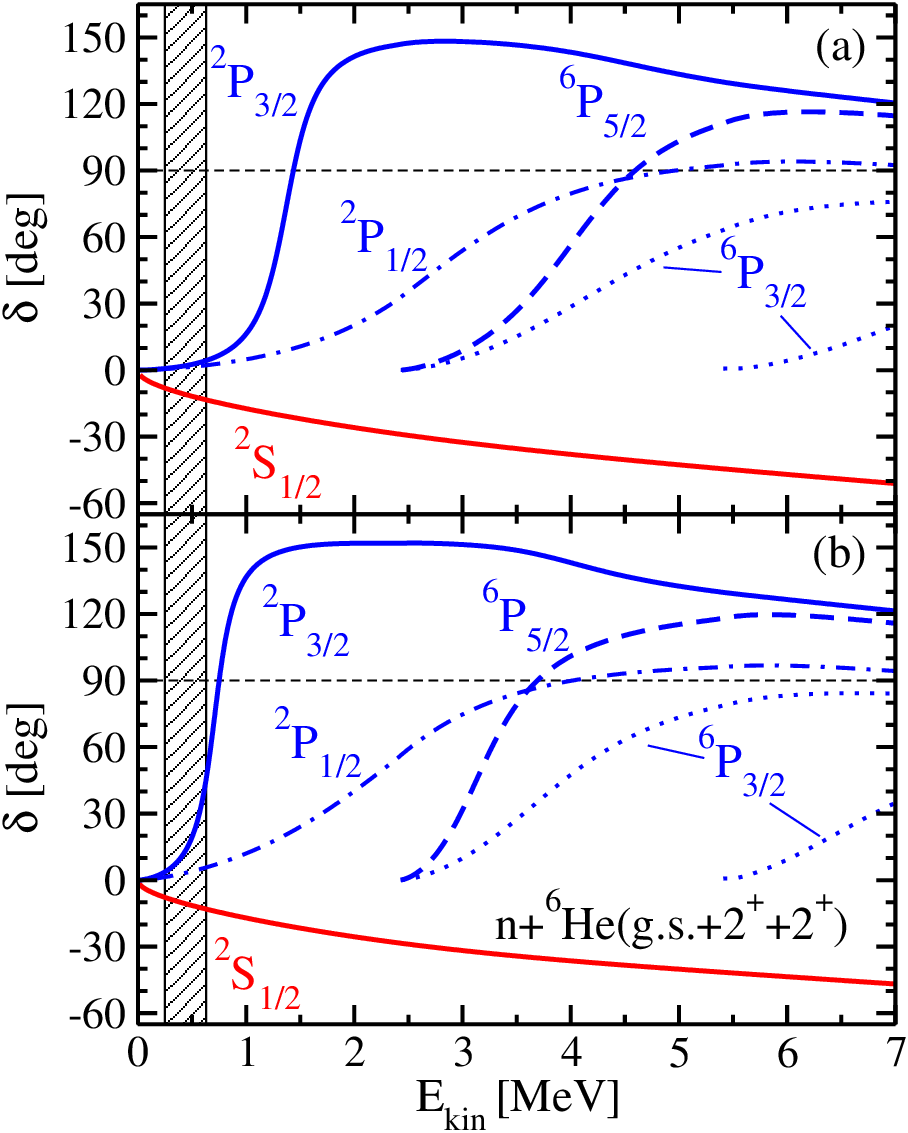}
\end{center}
\caption{(color online). NCSM/RGM (a) and NCSMC (b) $\elemA{6}{He}+n$ diagonal phase shifts (except $^6P_{3/2}$, which are eigenphase shifts)
as a function of the kinetic energy in the center of mass.
The dashed vertical area centered at $0.43 \mev$ indicates the experimental
centroid and width of the $\elemA{7}{He}$ ground state \cite{Stokes_1967,Cao2012}.
In all calculations the lowest three $\elemA{6}{He}$ states
have been included in the binary-cluster basis. The SRG-N$^3$LO $NN$ potential with $\Lambda=2.02$ fm$^{-1}$ within the $N_{\rm max}=12$ basis size and the HO frequency of $\hbar\Omega{=}16$ MeV were used. See text for further details.}
\label{6He_n_phase_shifts}
\end{figure}
The NCSM/RGM and NCSMC phase shifts for the $n+\elemA{6}{He}$ five $P$-wave 
and the 
$^2S_{1/2}$ channels are shown in Fig.~\ref{6He_n_phase_shifts}. All curves 
have been obtained 
including the three lowest $\elemA{6}{He}$ states (i.e.,
the $0^+$ ground state and the two lowest $2^+$ excited states) within the $N_{\rm max}=12$ HO basis. 
The model space of the NCSMC calculations [panel (b)] additionally includes 
ten $\elemA{7}{He}$ NCSM eigenstates, as described 
at the beginning of this section.
The dashed vertical area centered at $0.43 \mev$ indicates the experimental centroid and width of the $\elemA{7}{He}$ ground state~\cite{Stokes_1967,Cao2012}.
As expected 
from a variational calculation, the introduction of the additional square-integrable $A$-body
basis states $\ket{A \lambda J^\pi T}$ [{\em i.e.}, going from panel (a) to panel (b) of Fig.~\ref{6He_n_phase_shifts}] 
lowers the centroid values of all $\elemA{7}{He}$ resonances. 
In particular, the $3/2^-$ ground 
and $5/2^-$ excited states of $\elemA{7}{He}$ are pushed toward 
the $\elemA{6}{He}+n$ threshold, closer to their respective experimental positions. The resonance widths also shrink toward the observed data as we discuss below. We note that we also calculated higher partial waves, {\em e.g.} 
$D$-waves, in both approaches. However, the corresponding phase shifts are very small and do not present any 
interesting structures in the energy range displayed in Fig.~\ref{6He_n_phase_shifts}. Therefore, we did not include 
them in the figure. Unlike the $P$-wave resonances, the 
influence of the $^7$He positive-parity NCSM eigenstates on these phase shifts is rather weak. 

\begin{figure}
\begin{center}
\includegraphics[clip=,width=0.45\textwidth]{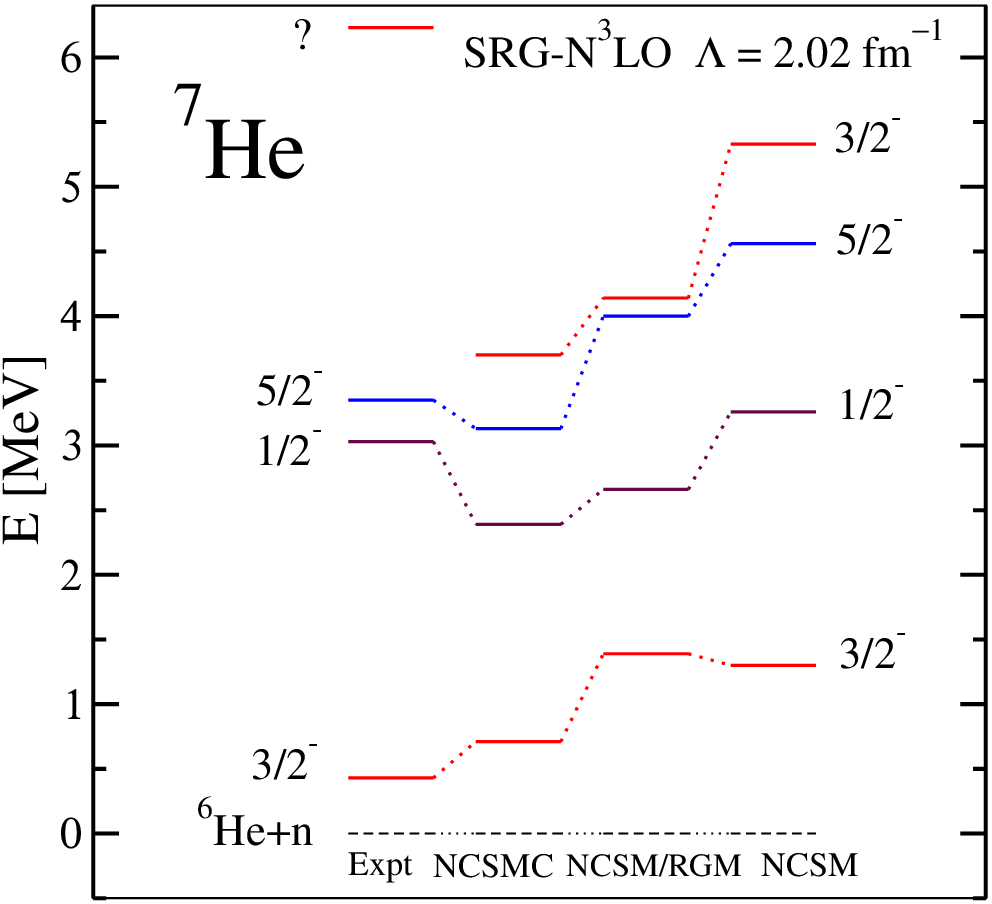}
\end{center}
\caption{(color online). Experimental and theoretical centroid energies for $\elemA{7}{He}$
resonances, with the $\elemA{6}{He} + n$ threshold as the energy reference. 
The experimental energy of the $1/2^-$ resonance is taken from Ref.~\cite{Wuosmaa_2005}.
The theoretical values for NCSM/RGM and NCSMC 
correspond to the $n{-}^6$He kinetic energy in the center of mass when the derivative of the phase shift is maximal, see text for details. The information on the width of the states is given in Table~\ref{6He_n_table_widths}. The calculations are carried out as described in Tab.~\ref{tab:NCSM_He_gs}, Fig.~\ref{6He_n_phase_shifts} and in the text.}
\label{6He_n_energies}
\end{figure}
The experimental centroid of the accepted $3/2^-$ and $5/2^-$ resonances in $\elemA{7}{He}$ as well as the possible $1/2^-$ state at $3.03 \mev$~\cite{Wuosmaa_2005} are shown in Fig.~\ref{6He_n_energies} together with our $N_{\rm max}=12$ predictions. 
For NCSM/RGM and NCSMC, the resonance centroids are calculated as the values of the kinetic energy in the center of mass $E_{kin}$ for which the first derivative of the phase shifts  is maximal~\cite{Thompson_priv}. 
The resonance widths are subsequently computed from the phase shifts according to (see, e.g., Ref.~\cite{Thompson}): 
\begin{equation}\label{eq:6Hen_10}
  \Gamma=\left. \frac{2}{{\rm d} \delta(E_{kin}) / {\rm d} E_{kin}}\right|_{E_{kin}=E_R}\,,
\end{equation}
where $E_R$ is the resonance centroid, evaluated as discussed above, and the phase shift are expressed in radians. 
Computed widths and $E_R$ values are reported in Table~\ref{6He_n_table_widths}, together with the available experimental data. An alternative, though less general, choice for the resonance energy could be the kinetic energy corresponding to a phase shift of $\pi/2$ (dashed horizontal lines in Fig.~\ref{6He_n_phase_shifts}). 
While the procedure of Eq.~(\ref{eq:6Hen_10}) is safely applicable to sharp resonances, broad resonances
would in principle require an analysis of the scattering matrix in the complex plane. 
Here, we are more interested in a qualitative discussion of the results, and will use the above extraction procedure for broad resonances as well.
Though the two alternative ways of 
choosing $E_R$ 
lead to basically identical results for our calculated $3/2^-_1$ resonance, the same is not true for the broader $5/2^-$ resonance and the very broad $1/2^-$ resonance. 
The less general $\pi/2$ condition, which is not valid for broad resonances, would result in $E_R\sim 3.7$ MeV and $\Gamma\sim 2.4$ MeV for the $5/2^-$ resonance and $E_R\sim 4$ MeV (see Fig.~\ref{6He_n_phase_shifts}) and $\Gamma\sim 13$ MeV for the $1/2^-$ resonance. 
\begin{table}[t]
\begin{center}
\begin{ruledtabular}
\begin{tabular}{c|ccc|cc|cc|c}
$J^\pi$        & \multicolumn{3}{c|}{experiment} & \multicolumn{2}{c|}{NCSMC} & \multicolumn{2}{c|}{NCSM/RGM} & NCSM \\ 
               &   $E_R$   & $\Gamma$ & Ref.                     &   $E_R$  & $\Gamma$   &   $E_R$ & $\Gamma$ & $E_R$   \\
\hline
$3/2^-$        & 0.430(3)& 0.182(5)& \cite{Cao2012}     &  0.71 & 0.30  & 1.39  & 0.46       & 1.30 \\
$5/2^-$        & 3.35(10) & 1.99(17) & \cite{Tilley2002} &  3.13 & 1.07 & 4.00  & 1.75        & 4.56 \\
$1/2^-$        & 3.03(10)  & 2     & \cite{Wuosmaa_2005}   &  2.39 & 2.89 & 2.66  & 3.02        & 3.26 \\
                     & 3.53    & 10    & \cite{Boutachkov_2005} &   &   &       &   &   \\
                     & 1.0(1)     & 0.75(8)  & \cite{Meister_2002} &   &   &       &   &   \\
\end{tabular} 
\caption{Experimental and theoretical values for the resonance centroids and widths in MeV for the
$3/2^-$ ground state and the $5/2^-$ and $1/2^-$ excited states of $\elemA{7}{He}$. Calculations are carried out as described in Tab.~\ref{tab:NCSM_He_gs}, Fig.~\ref{6He_n_phase_shifts} and in the text.}
\label{6He_n_table_widths}
\end{ruledtabular}
\end{center}
\end{table}

Interestingly, the NCSM eigenenergy for the $3/2^-_1$ ground state
resonance is close to the energy centroid found within the NCSM/RGM approach. This is accidental as both calculations are deficient in different ways. The NCSM lacks the description of long-range correlations due to the HO basis truncation, while the NCSM/RGM lacks a proper description of short- and medium-range correlations due to the omission of higher excited states of the $^6$He target. In the NCSMC, a significant energy shift is brought by the coupling of the two basis, with a quenching 
of the separation energy by almost $0.7 \mev$, closer to the experimental findings. At the same time, in the NCSMC calculation, the resonances become sharper, with narrower widths, 
once again in a better agreement with experiment. 
Our NCSMC $3/2^-$ g.s.\ resonance position and width slightly overestimate measurement (e.g., the latest determination from the recoil proton tagged knockout reaction for $^8$He~\cite{Cao2012} finds $E_R=0.430(3)$ MeV and $\Gamma=0.182(5)$ MeV). At the same time, predictions for the $5/2^-$ resonance are lower compared to experiment~\cite{Korsheninnikov_1998,Tilley2002}, though our determination of the width should be take with some caution in this case.

In all three approaches considered here, the $1/2^-$ resonance is predicted below the $5/2^-$ excited state. At the same time one has to keep in mind that the NCSM approach is not expected to provide a reliable
description of broad resonances 
and 
that our determination of the $1/2^-$ resonance position in the NCSM/RGM and NCSMC has to be taken with some caution, as explained above. 
From an experimental standpoint, the situation concerning the $1/2^-$ resonance is not clear as discussed in the beginning of this section and documented in Table~\ref{6He_n_table_widths}. While the centroid energies determined in the experiments of Refs.~\cite{Wuosmaa_2005, Wuosmaa_2008}  and~\cite{Boutachkov_2005} are comparable, the widths are very different. Within the present determination of $E_R$ and $\Gamma$, the NCSMC results are in fair agreement with the $1/2^-$ properties measured in the neutron pick-up and proton-removal reactions experiments of Refs.~\cite{Wuosmaa_2005} and~\cite{Wuosmaa_2008}. Our calculations definitely do not support the hypothesis of a low-lying ($E_R{\sim} 1$ MeV) narrow ($\Gamma \leq 1$ MeV) $1/2^-$ resonance~\cite{Markenroth_2001,Meister_2002,Skaza_2006,Ryezayeva_2006,Lapoux_2006}.

\begin{figure}[t]
\begin{center}
\includegraphics[clip=,width=0.45\textwidth]{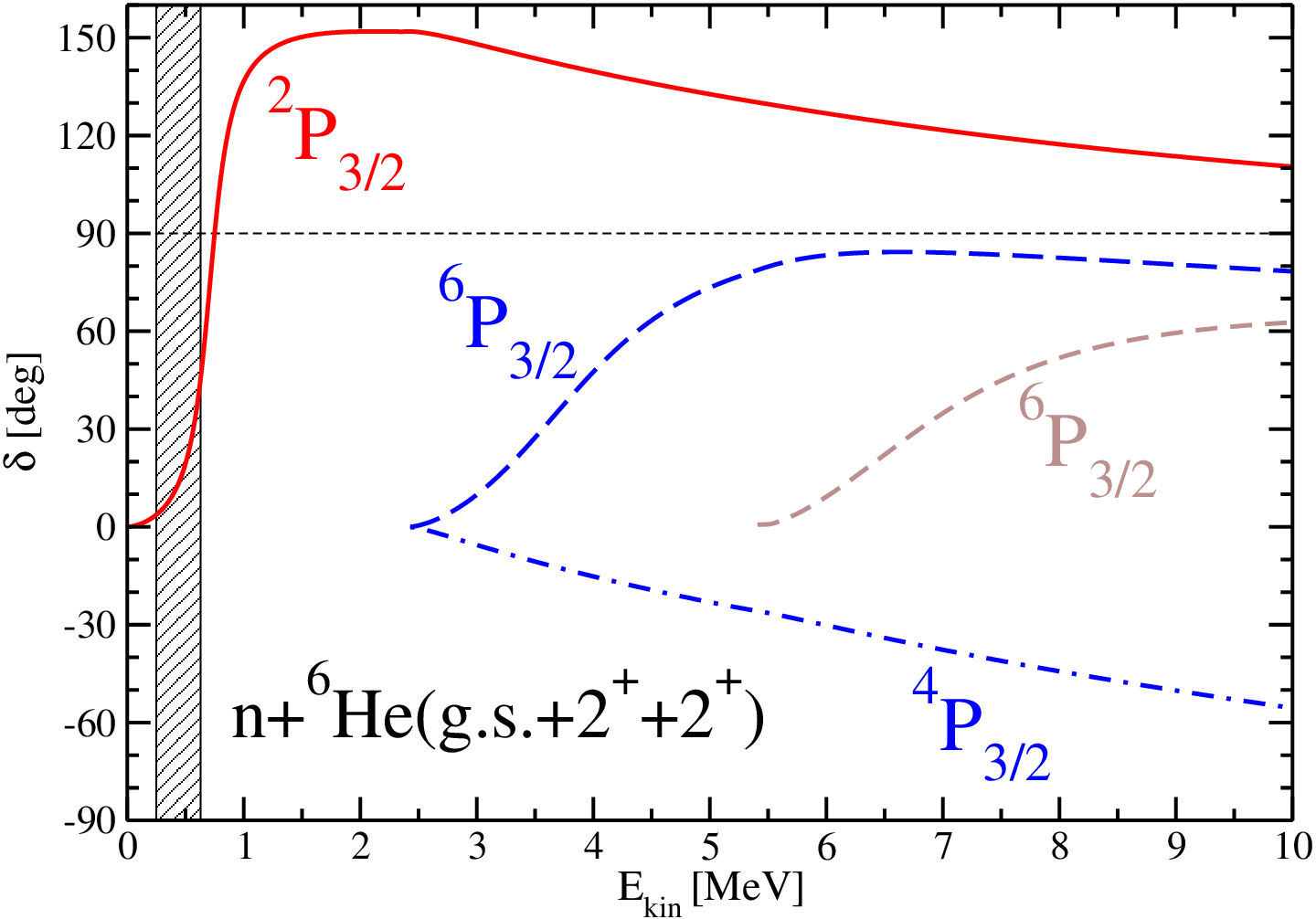}
\end{center}
\caption{(color online). NCSMC $\elemA{6}{He}+n$ $3/2^-$ $P$-wave eigenphase shifts as a function of the kinetic energy in the center of mass.
Calculations are carried out as described in Fig.~\ref{6He_n_phase_shifts} (b). See text for further details.}
\label{6He_n_eigenphase_shifts}
\end{figure}
\begin{figure}[!ht]
\begin{center}
\includegraphics[clip=,width=0.45\textwidth]{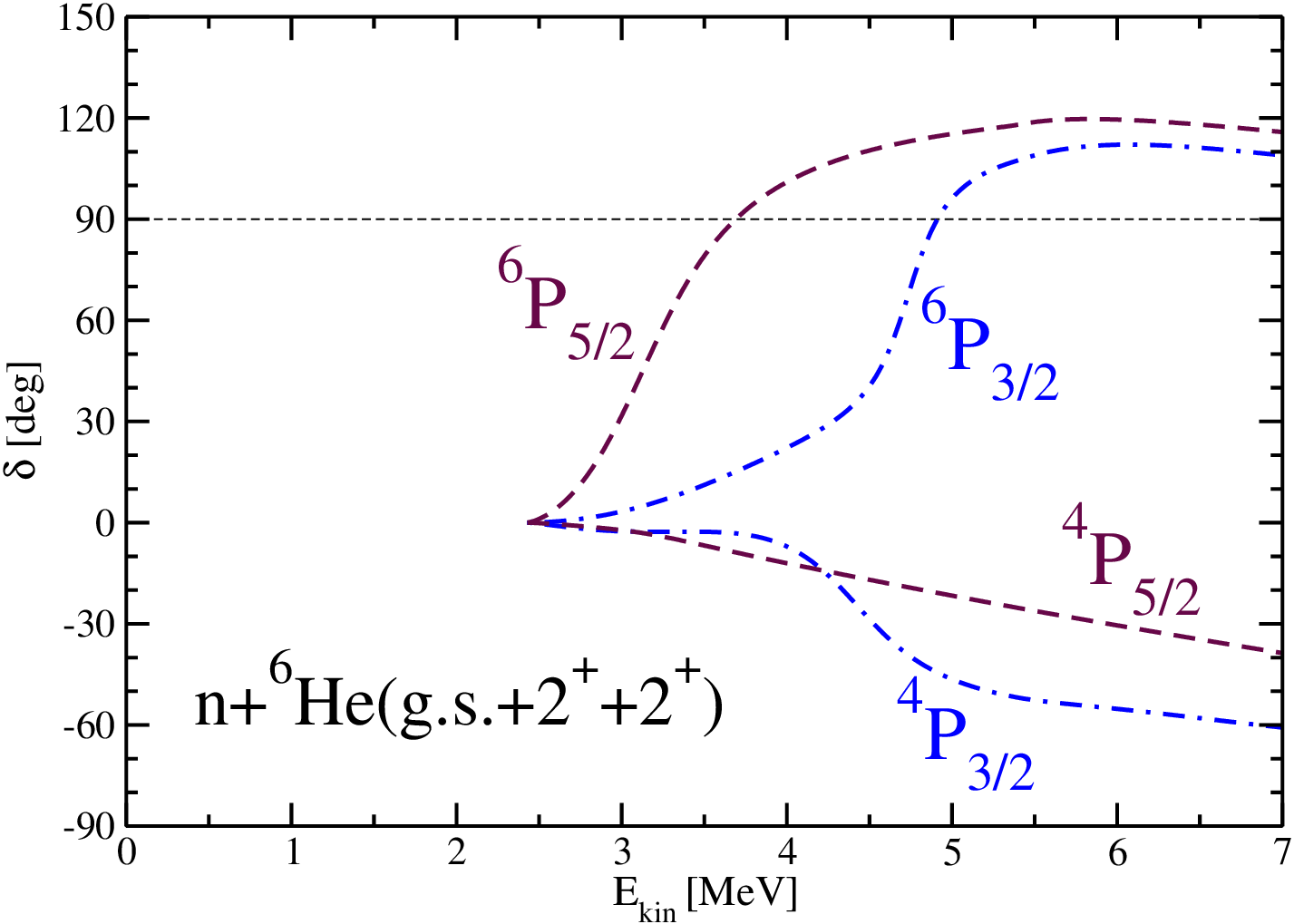}
\end{center}
\caption{(color online). NCSMC $\elemA{6}{He}+n$ $3/2^-_2$ and $5/2^-$ diagonal $P$-wave phase shifts as a function of the kinetic energy in the center of mass. The calculation is carried out as described in Fig.~\ref{6He_n_phase_shifts} (b). See text for further details.}
\label{fig_Pwaves}
\end{figure}

We also note that our NCSMC calculations predict two broad $^6P_{3/2}$ resonances (dominated, respectively, by the first and second $2^+$ states of $^6$He) at about 3.7 MeV and 6.5 MeV with widths of 2.8 MeV and 4.3 MeV, respectively. As shown in Figs.~\ref{6He_n_phase_shifts} and~\ref{6He_n_eigenphase_shifts}, the corresponding eigenphase shifts do not cross $\pi/2$. In Fig.~\ref{6He_n_eigenphase_shifts}, we present all $P$-wave eigenphase shifts in a broader energy range up to 10 MeV. There is a considerable mixing of the $P$ waves around the $3/2^-_2$ resonance as it can be seen by comparing the eigenphase shifts of Fig.~\ref{6He_n_eigenphase_shifts} with the diagonal $^6P_{3/2}$ and $^4P_{3/2}$ phase shifts of Fig.~\ref{fig_Pwaves}. The mixing parameter for other resonances is very small (of course, there is no mixing below the $n+^6$He$(2^+_1)$ state threshold).  In experiment, there is a resonance of undetermined spin and parity at 6.2(3) MeV with a width of 4(1) MeV~\cite{Tilley2002}.

The level order predicted in other theoretical calculations mostly agrees with our present findings~\cite{GFMC,Wuosmaa_2005,Myo2007}. The widths of the $^7$He states were calculated recently in a $^4$He$+n+n+n$ cluster model~\cite{Myo2007}. The $1/2^-$ state was found at low excitation energy (${\sim}1.05$ MeV), but with a width of 2.19 MeV, {\em i.e.}, close to what we find. The width of the $5/2^-$ resonance, 1.5 MeV, obtained in Ref.~\cite{Myo2007} is also comparable to our prediction. Two $3/2^-$ resonances in addition to the g.s.\ resonance were reported in Ref.~\cite{Myo2007}. One of them just above the $5/2^-$ state with a width of 1.95 MeV, while the other at the excitation energy of about 5.3 MeV and a width of 5.77 MeV. This is qualitatively similar to our results although, in our case, the $3/2^-_2$ resonance is broader by 0.85 MeV.

The $^7$He resonances were also investigated in RGM calculations of Ref.~\cite{Wurzer1997} using a semi-realistic $NN$ potential. The 
ordering of the resonances found in this study is the same as in ours and the $^2P_{3/2}$ g.s.\ resonance phase shift is also in close agreement with our results. On the other hand, unlike in our calculations, the $1/2^-$ and $5/2^-$ $P$-wave resonances of Ref.~\cite{Wurzer1997} do not cross $\pi/2$. While the $^6P_{3/2}$ and $^4P_{3/2}$ diagonal phase shifts qualitatively agree with ours, interestingly, the $5/2^-$ resonance appears in the $^4P_{5/2}$ partial wave rather than in the $^6P_{5/2}$, as found in our calculations. 
The $5/2^-$ $P$-waves are reversed in Ref.~\cite{Wurzer1997} compared to our calculations (see Fig.~\ref{fig_Pwaves} for the $^6P_{3/2}$, $^4P_{3/2}$, $^6P_{5/2}$, and $^4P_{5/2}$ diagonal phase shifts). 

The Helium isotope g.s.\ properties, including those of $^7$He, were also recently investigated within the complex coupled-cluster method~\cite{Hagen2007}. Using a realistic low-momentum $NN$ interaction, the coupled-cluster singles and doubles (CCSD) calculations underbinded substantially the ground states of $^{3-10}$He compared to experiment. However, they correctly predicted $^5$He and $^7$He unstable with respect to neutron emission. The width of the $^7$He ground state resonance, 0.26 MeV 
is quite close to that calculated here.

Finally, we note that the NCSMC g.s.\ resonance energy, 0.71 MeV, is lower but still compatible with the extrapolated NCSM value of 0.98(29) MeV (see Tables~\ref{tab:NCSM_He_gs} and \ref{6He_n_table_widths}).

\section{Conclusions and outlook}
\label{concl}

We introduced a new unified approach to nuclear bound and continuum states based on the coupling of a square-integrable basis ($A$-body NCSM eigenstates), suitable for the description of many-body correlations, and a continuous basis (NCSM/RGM cluster states) suitable for a description of long-range correlations, cluster correlations and scattering. This {\it ab initio} method, which we call no-core shell model with continuum, 
is capable of describing efficiently: $i)$ short- and medium-range nucleon-nucleon correlations thanks 
to the large HO basis expansions used to obtain the NCSM eigenstates, and 2) 
long-range cluster correlations thanks to the NCSM/RGM cluster-basis expansion. As a consequence, its convergence properties are superior to either NCSM or NCSM/RGM. 

We demonstrated the 
potential of the NCSMC in calculations of $^7$He resonances. Starting from a realistic soft SRG-N$^3$LO $NN$ potential that describes accurately two-nucleon 
properties and, with the 
choice of $\Lambda=2.02$ fm$^{-1}$ for the 
SRG evolution parameter, also predicts $^3$H and $^4$He binding energies close to experiment, we calculated $^6$He and $^7$He eigenstates in the NCSM and used them as input to the coupled-channel NCSMC equations. We found the $^6$He g.s.\ energy in very good agreement with experiment. The results for the $3/2^-$ g.s.\ resonance as well as for the well-established $5/2^-$ resonance of $^7$He
are in reasonable agreement with experiment. Our results for the controversial $1/2^-$ resonance are in fair agreement with the neutron pick-up and proton-removal reactions experiments of Refs.~\cite{Wuosmaa_2005, Wuosmaa_2008}. Our calculations definitely do not support the hypothesis of a low lying ($E_R{\sim} 1$ MeV) narrow ($\Gamma \leq 1$ MeV) $1/2^-$ resonance. We also predict two broad currently unobserved $^6P_{3/2}$ resonances at about 3.7 MeV and 6.5 MeV, respectively.

The NCSMC calculations do not involve any adjustable parameter except for those used in the construction of the input $NN$ (or three-nucleon) potentials. Computations depend on the size of the HO basis, 
the HO frequency, and the number of eigenstates included in the model space. 
We investigate the convergence behavior of the approach with respects to these expansions. Due to the over-completeness of the NCSMC basis, the convergence rate is superior to that achievable with either NCSM or NCSM/RGM. The 
advantages of the NCSMC are expected to become even more evident 
in calculations with composite projectiles (such as deuteron, $^3$H, or $^3$He) that require the use of a large number of pseudostates in the NCSM/RGM (or other cluster-based approaches) to account for virtual breakup effects. The contribution of 
the pseudostates is expected to be suppressed 
in the NCSMC approach. Extension of the NCSMC formalism to the case of composite projectiles, the inclusion of three-nucleon interactions, and the coupling of three-body clusters are under way.

\acknowledgments

Computing support for this work came in part from the LLNL institutional Computing Grand Challenge program. This work was prepared in part by the LLNL under Contract DE-AC52-07NA27344. Support from the U.\ S.\ DOE/SC/NP (Work Proposal No.\ SCW1158) and the Natural Sciences and Engineering Research Council of Canada (NSERC) Grant No. 401945-2011, and from the U.\ S.\ Department of Energy Grant DE-FC02-07ER41457 is acknowledged. TRIUMF receives funding via a contribution through the National Research Council Canada.
This research was supported in part by the PAI-P6-23 of the Belgian Office for Scientific Policy
and by the European Union Seventh Framework Programme under grant agreement No. 62010.
\begin{widetext}
\appendix
\section{}\label{Appendix_A}
In this appendix we briefly outline the explicit steps for the derivation of the  orthogonalized cluster form factors of Eq.~(\ref{eq:formalism_150}) and~(\ref{eq:formalism_160}) and provide their algebraic expressions.

The orthogonalized cluster form factor in $r$-space representation of Eq.~(\ref{eq:formalism_150}) reads
\begin{align}\label{eq:formalism_ff_2}
             \bar{g}_{\lambda\nu}(r)  & =  \sum_{\nu'}\int dr' {r'}^2 
                               \braket{A \lambda J^\pi T} 
                                      {\hat{\mathcal{A}}_{\nu'} \Phi_{\nu' r'}^{J^\pi T  }}
                               \; \mathcal{N}_{\nu' \nu}^{-\frac{1}{2}}(r',r) 
                 \nonumber \\
                         &=   \sum_{n\in P}R_{n \ell}(r) \sum_{\nu' n'\in P}
                                     \braket{A \lambda J^\pi T}
                                            {\hat{\mathcal{A}}_{\nu'} \Phi_{\nu' n'}^{J^\pi T  }}
                                     \; \mathcal{N}_{\nu' n', \nu n}^{-\frac{1}{2}} \\
                        &= \sum_{n\in P}R_{n \ell}(r)\; \bar{g}_{\lambda \nu n} \; ,
\end{align}
where the orthogonalized cluster form factor in the  model-space is given by the model-space non-orthogonalized cluster form factor times the model-space norm kernel:
\begin{eqnarray}\label{eq:formalism_ff_2.2}
  \bar{g}_{\lambda \nu n} & = & \sum_{\nu' n'\in P}
                                     \braket{A \lambda J^\pi T}
                                            {\hat{\mathcal{A}}_{\nu'} \Phi_{\nu' n'}^{J^\pi T  }}
                                     \; \mathcal{N}_{\nu' n', \nu n}^{-\frac{1}{2}} 
                       =  
                       			\sum_{\nu' n'\in P}
                                     g_{\lambda \nu' n'}
                                     \; \mathcal{N}_{\nu' n', \nu n}^{-\frac{1}{2}}\,.
\end{eqnarray}
At the same time, the translational-invariant non-orthogonalized cluster form factors in the model space, $g_{\lambda \nu n}$, can be conveniently derived starting from the Slater-determinant (SD) NCSM eigenstates,
\begin{align}
	\ket{A \lambda J^\pi T}_{SD} = \ket{A \lambda J^\pi T} \,\varphi_{00}(\vec{R}^{(A)}_{c.m.})\,, \label{SD-eigenstate}
\end{align}
and the SD channel states
\begin{align}
	\ket{\Phi_{\nu n}^{J^\pi T}}_{SD} = &
	\Big[ \left(
         \ket{A-a \; \alpha_1 I_1^{\pi_1}T_1}_{SD}\ket{a \; \alpha_2 I_2^{\pi_2}T_2}
         \right)^{(sT)} 
         Y_\ell(\hat{R}^{(a)}_{c.m.})
         \Big]^{(J^{\pi}T)} R_{n\ell}(R^{(a)}_{c.m.})\,,\label{SD-channel}
\end{align}
and removing the spurious motion of the center of mass. Here, the c.m.\ coordinates of Eqs.~(\ref{SD-eigenstate}) and (\ref{SD-channel}) are given by 
\begin{align}
	\vec{R}^{(A)}_{c.m.} = \frac{1}{\sqrt A} \sum_{i=1}^A \vec{r_i}\,, & \qquad \vec{R}^{(a)}_{c.m.} = \frac{1}{\sqrt a} \sum_{i=A-a+1}^A \vec{r_i}\,,
\end{align}
and $\varphi_{00}(\vec{R}^{(A)}_{c.m.})$ is the HO wave function $R_{00}(R^{(A)}_{c.m.}) Y_{00}(\hat{R}^{(A)}_{c.m.})$.  The resulting expression for the non-orthogonalized cluster form factor in the single-nucleon projectile ($a=1$) basis is:
\begin{eqnarray}\label{eq:formalism_ff_2.4}
  g_{\lambda \nu n} & = & \braket{A \lambda J^\pi T}
                               {\hat{\mathcal{A}}_{\nu} \Phi_{\nu n}^{J^\pi T  }} \nonumber \\
                              & = &  \frac{1}{  \braket{n\ell00,\ell}{00n\ell,\ell}_{\frac{1}{(A-1)}} } \;{}_{SD}\braket{A \lambda J^\pi T}
                               {\hat{\mathcal{A}}_{\nu} \Phi_{\nu n}^{J^\pi T  }}_{SD} \nonumber \\
                  & = & \frac{1}{  \braket{n\ell00,\ell}{00n\ell,\ell}_{\frac{1}{(A-1)}} }
                        \frac{1}{\hat{J}\hat{T}}\sum_j (-1)^{I_1 + J + j}\hat{s} \hat{j} 
                               \left\{
			         \begin{array}{ccc}
				   I_1 & \tfrac12 & s \\
				   \ell   & J   & j \\
				 \end{array}
			       \right\}
                                 {}_{SD}\bra{A \lambda J^\pi T}||a^\dagger_{n\ell j\frac{1}{2}}||
                                 \ket{A-1 \alpha_1 I_1^{\pi_1} T_1}_{SD} \; .
\end{eqnarray}
The Moshinsky brackets $\braket{n\ell00,\ell}{00n\ell,\ell}$ allows us to transform from the SD to the Jacobi-coordinate
states. This expression was first derived in Ref.~\cite{Navratil:2004tw} where further details on the derivation can be found.

The orthogonalized coupling form factor in $r$-space representation of Eq.~(\ref{eq:formalism_160}) reads
\begin{align}\label{eq:formalism_ff_4} 
            \bar{h}_{\lambda\nu}(r)  &=  \sum_{\nu'}\int dr' {r'}^2 
                               \matrEL{A \lambda J^\pi T}
                                      {\hat{H}}
                                      {\hat{\mathcal{A}}_{\nu'} \Phi_{\nu' r'}^{J^\pi T}} \; \mathcal{N}_{\nu' \nu}^{-\frac{1}{2}}(r',r) \nonumber \\
                        & =   \sum_{n\in P}R_{n \ell}(r) \sum_{\nu' n'\in P}
                                     \matrEL{A \lambda J^\pi T}
                                            {\hat{H}}
                                            {\hat{\mathcal{A}}_{\nu'} \Phi_{\nu' n'}^{J^\pi T}}
                                     \; \mathcal{N}_{\nu' n', \nu n}^{-\frac{1}{2}} 
                        + R_{n_{max}+1\, \ell}(r) \braket{A \lambda J^\pi T}
                                                        {\hat{\mathcal{A}}_{\nu}\Phi_{\nu n_{max}}^{J^\pi T}}
                                                        \matrEL{n_{max} \ell}{\hat{T}_{rel}}{n_{max}+1\, \ell}
                         \nonumber \\
                        & =  \sum_{n\in P}R_{n \ell}(r) \bar{h}_{\lambda \nu n} 
                         +  
                              R_{n_{max}+1\, \ell}(r)\;  \matrEL{n_{max} \ell}{\hat{T}_{rel}}{n_{max}+1\, \ell} \; g_{\lambda \nu n_{max}} \; ,
\end{align}
where
\begin{eqnarray}\label{eq:formalism_ff_4.2}
  \bar{h}_{\lambda \nu n} & = & \sum_{\nu' n'\in P}
                                     \matrEL{A \lambda J^\pi T}
                                            {\bar{H}}
                                            {\hat{\mathcal{A}}_{\nu'} \Phi_{\nu' n'}^{J^\pi T  }}
                                     \mathcal{N}_{\nu' n', \nu n}^{-\frac{1}{2}} 
                        \equiv  
                              \sum_{\nu' n'\in P}
                                     h_{\lambda \nu' n'}
                                     \mathcal{N}_{\nu' n', \nu n}^{-\frac{1}{2}}
\end{eqnarray}
is the orthogonalized coupling form factor in the model space. In deriving the above expression, one has to pay attention in taking into account the contribution of transitions to basis states outside of the model space brought about by the relative kinetic-energy operator. The model-space non-orthogonalized coupling form factor $h_{\lambda \nu n}$ can be derived in a similar fashion as Eq.~(\ref{eq:formalism_ff_4}), and is given by:
\begin{align}\label{eq:formalism_ff_4.4}
             h_{\lambda \nu n}  &=  \matrEL{A \lambda J^\pi T}
                               {\hat{H}}
                               {\hat{\mathcal{A}}_{\nu} \Phi_{\nu n}^{J^\pi T  }} \nonumber\\[2mm]
                  &=   \sum_{\nu'}\sum_{n' \in P} g_{\lambda \nu' n'} \;
        \matrEL{n'\ell'}{\hat{T}_{rel}}{n\ell} + g_{\lambda \nu n} \; E_\nu 
                  \nonumber \\
  & \phantom{=} +   \frac{1}{ \braket{n\ell00,\ell}{00n\ell,\ell}_{\frac{1}{(A-1)}} }
         \sum_j (-1)^{I_1 + J - j}\hat{s} 
                               \left\{
			         \begin{array}{ccc}
				   I_1 & \tfrac12 & s \\
				   \ell   & J   & j \\
				 \end{array}
			       \right\}
                               \frac{1}{2\sqrt{2}} \nonumber \\[2mm]
  &  \phantom{=}\times \sum_{J'T'}\sum_{(nlj)_{abc}}\frac{\hat{J'}\hat{T}'}{\hat{J}\hat{T}} 
      \sqrt{1+\delta_{n_a l_a j_a, n_b l_b j_b}} \sqrt{1+\delta_{n_c l_c j_c, n \ell j}} \nonumber \\[4mm]
  &  \phantom{=}\times
      \matrEL{(n_a l_a j_a \tfrac{1}{2}, n_b l_b j_b \tfrac{1}{2}) J' T'}
             {V}
             {(n_c l_c j_c \tfrac{1}{2}, n\ell j \tfrac{1}{2}) J' T'} \nonumber \\[4mm]
  &  \phantom{=}\times                       
       {}_{SD}\matrEL{A \lambda J^\pi T}
              {|| ( ( a^\dagger_{n_a l_a j_a \frac{1}{2}}  a^\dagger_{n_b l_b j_b \frac{1}{2}} )^{(J'T')} 
                \tilde{a}_{n_c l_c j_c \frac{1}{2}} )^{(j\frac{1}{2})}||}
              {(A-1)\alpha_1 I_1 T_1}_{SD} \; .
\end{align}
We note that the point-Coulomb contribution introduced in Eq.~(\ref{eq:formalism_40}) is omitted in the above expressions for simplicity. It is zero in the present application to $^7$He. Finally, $E_\nu$ is the sum of the eigenergies of the two clusters. 

\end{widetext}

\bibliographystyle{apsrev}

\end{document}